\documentclass[12pt,a4paper]{article}
\usepackage[british]{babel}
\usepackage{epsfig}
\begin{document}
\date{}

\title{
{\vspace{-20mm} \normalsize
\hfill \parbox[t]{50mm}{DESY 03-072  \\
                        MS-TP-03-05}}\\[20mm]
 Quark mass dependence of masses and decay \\
 constants of the pseudo-Goldstone bosons in QCD}

\author{qq+q Collaboration                                  \\[0.5em]
        F. Farchioni                                        \\[0.5em]
        Westf\"alische Wilhelms-Universit\"at M\"unster,    \\
        Institut f\"ur Theoretische Physik,                 \\
        Wilhelm-Klemm-Strasse 9, D-48149 M\"unster, Germany \\[1em]
        I. Montvay, E. Scholz, L. Scorzato                  \\[0.5em]
        Deutsches Elektronen-Synchrotron DESY               \\
        Notkestr.\,85, D-22603 Hamburg, Germany}

\newcommand{\be}{\begin{equation}}                                              
\newcommand{\ee}{\end{equation}}                                                
\newcommand{\half}{\frac{1}{2}}                                                 
\newcommand{\rar}{\rightarrow}                                                  
\newcommand{\lar}{\leftarrow}
\newcommand{\LCB}{\raisebox{-0.3ex}{\mbox{\LARGE$\left\{\right.$}}}
\newcommand{\RCB}{\raisebox{-0.3ex}{\mbox{\LARGE$\left.\right\}$}}}
\newcommand{\U}{\mathrm{U}}
\newcommand{\SU}{\mathrm{SU}}
\newcommand{\bteq}[1]{\boldmath$#1$\unboldmath}


\maketitle

\abstract{
 The dependence of the pseudoscalar meson masses and decay constants on
 sea and valence quark masses is compared to next-to-leading order (NLO)
 Chiral Perturbation Theory (ChPT).
 The numerical simulations with two light dynamical quark flavors are
 performed with the Wilson-quark lattice action at gauge coupling
 $\beta=5.1$ and hopping parameters $\kappa=0.176,\; 0.1765,\; 0.177$
 on a $16^4$ lattice.
 ${\cal O}(a)$ lattice artifacts are taken into account by applying
 chiral perturbation theory for the Wilson lattice action.
 The values of the relevant combinations of Gasser-Leutwyler constants
 $L_4$, $L_5$, $L_6$ and $L_8$ are estimated.}

\section{Introduction}\label{sec1}

 The low energy dynamics of strong interactions in the pseudo-Goldstone
 boson sector of QCD is constrained by the non-linear realization of
 spontaneously broken chiral symmetry \cite{WEINBERG}.
 In an expansion in powers of momenta and light quark masses a few low
 energy constants -- the Gasser-Leutwyler constants -- appear which
 parameterize the strength of interactions in the low energy chiral
 Lagrangian \cite{CHPT}.
 The Gasser-Leutwyler constants are free parameters which can be
 constrained by analyzing experimental data.
 In the framework of lattice regularization they can be determined
 from first principles by numerical simulations.
 In experiments one can investigate processes with different momenta
 but the quark masses are, of course, fixed by Nature.
 In numerical simulations there is, in principle, much more freedom
 because, besides the possibility of changing momenta, one can also
 freely change the masses of the quarks.
 This allows for a precise determination of the Gasser-Leutwyler
 constants -- once the simulations reach high precision.
 First steps towards this goal have recently been done by several
 authors \cite{ALPHA:CHPT,UKQCD:CHPT,STAGGERED1,STAGGERED2} including
 our Collaboration \cite{NF2TEST,PRICE,VALENCE}.

 The main difficulty for numerical simulations in lattice QCD is to
 reach the regime of light quark masses where ChPT is applicable.
 The reason is the critical slowing down of simulation algorithms
 for small quark masses and lattice spacings.
 We apply the two-step multi-boson (TSMB) algorithm \cite{TSMB} which
 allows to perform simulations with small quark masses within the
 range of applicability of next-to-leading order (NLO) ChPT
 \cite{NF2TEST,PRICE}.

 Another important aspect of investigating the quark mass dependence
 in numerical simulations is the possibility to use ChPT for the
 extrapolation of the results to the physical values of $u$- and
 $d$-quark masses which would be very difficult to reach otherwise.
 In fact, ChPT can be extended by changing the {\em valence quark
 masses} in quark propagators independently from the {\em sea quark
 masses} in virtual quark loops which are represented in the path
 integral by the {\em quark determinant}.
 In this way one arrives at Partially Quenched Chiral Perturbation
 Theory (PQChPT) \cite{BERNARD-GOLT,SHARPE,GOLT-LEUNG}.
 The freedom of changing valence and sea quark masses substantially
 contributes to the power of lattice QCD both in performing quark mass
 extrapolations and in determining the values of the Gasser-Leutwyler
 constants \cite{SHARPE-SHORESH}.

 For a fast convergence of numerical results to the continuum limit it
 is important to explicitly deal with the leading ${\cal O}(a)$ lattice
 artifacts.
 An often used method is the application of the ${\cal O}(a)$ improved
 lattice action \cite{CLOVER}. 
 We apply an alternative technique \cite{RUPAK-SHORESH} which in the
 pseudo-Goldstone boson sector is equivalent to the ${\cal O}(a)$
 improvement of the lattice action. 
 In this method the (unimproved) Wilson action is used in the Monte
 Carlo generation of gauge configurations and the ${\cal O}(a)$ effects
 are compensated in PQChPT itself.
 This means that we apply {\em chiral perturbation theory for the
 Wilson lattice action.}
 Our calculations showed that in practice this method gives results
 with good precision \cite{VALENCE}.

 The plan of this paper is as follows: in the next two sections we
 collect the NLO (PQ)ChPT formulas for ratios of pseudoscalar meson
 masses and decay constants.
 In section \ref{sec2} a discussion of the general form of the NNLO
 tree-graph corrections is also included.
 In section \ref{sec4} the results of numerical simulations is
 presented.
 The last section is devoted to a summary and discussion.

\section{Valence quark mass dependence}\label{sec2}

 In this paper we use the notations introduced in \cite{VALENCE} which
 slightly differ from those of ref.~\cite{SHARPE-SHORESH} and 
 \cite{RUPAK-SHORESH}.
 The dimensionless variables for quark masses and ${\cal O}(a)$
 lattice artifacts are denoted, respectively, by 
\be \label{eq01}
\chi_A \equiv \frac{2B_0 m_q}{f_0^2} \ , \hspace{3em}
\rho_A \equiv \frac{2W_0 ac_{SW}}{f_0^2} \ .
\ee
 Here $m_q$ is the quark mass, $a$ the lattice spacing, $B_0$ and $W_0$
 are parameters of dimension mass and (mass)$^3$, respectively, which
 appear in the leading order (LO) chiral effective Lagrangian, $c_{SW}$
 is the coefficient of the ${\cal O}(a)$ chiral symmetry breaking term
 and $f_0$ is the value of the pion decay constant at zero quark mass.
 (Its normalization is such that the physical value is
 $f_0 \simeq 93\, {\rm MeV}$.)
 For fixed sea quark mass $\chi_S$ the dependence of the pseudoscalar
 meson mass and decay constant on the valence quark mass $\chi_V$
 can be described by the variables
\be \label{eq02}
\xi   \equiv \frac{\chi_V}{\chi_S} \ , \hspace{3em}
\eta_S  \equiv \frac{\rho_S}{\chi_S} \ .
\ee
 For instance, in case of a number of $N_s$ equal mass sea quarks the
 ratios of decay constants are given by
\begin{eqnarray}
Rf_{VV} \hspace*{-1.5em}&& \equiv \frac{f_{VV}}{f_{SS}}
= 1 + 4(\xi-1)\chi_S L_{S5}
\nonumber
\\[0.5em] && \label{eq03}
- \frac{N_s\chi_S}{64\pi^2}(1+\xi+2\eta_S)\log\frac{1+\xi+2\eta_S}{2}
+ \frac{N_s\chi_S}{32\pi^2}(1+\eta_S)\log(1+\eta_S) \ ,
\end{eqnarray}
 and similarly
\begin{eqnarray}
Rf_{VS} \hspace*{-1.5em}&& \equiv \frac{f_{VS}}{f_{SS}}
= 1 + 2(\xi-1)\chi_S L_{S5} + \frac{\chi_S}{64N_s\pi^2}(\xi-1)
- \frac{\chi_S}{64N_s\pi^2}(1+\eta_S)\log\frac{\xi+\eta_S}{1+\eta_S}
\nonumber
\\[0.5em] && \label{eq04}
- \frac{N_s\chi_S}{128\pi^2}(1+\xi+2\eta_S)\log\frac{1+\xi+2\eta_S}{2}
+ \frac{N_s\chi_S}{64\pi^2}(1+\eta_S)\log(1+\eta_S) \ .
\end{eqnarray}
 $L_{Sk}$ ($k=5$) denotes the relevant Gasser-Leutwyler coefficient at
 the scale $f_0\sqrt{\chi_S}$.
 This is related to $\bar{L}_k$ defined at the scale $f_0$ and
 $L^\prime_k$ defined at the generic scale $\mu$ according to
\be \label{eq05}
L_{Sk} = \bar{L}_k - c_k\log(\chi_S) 
= L^\prime_k - c_k\log(\frac{f_0^2}{\mu^2}\chi_S) \ .
\ee
 with the constants $c_k,\; (k=4,5,6,8)$ given below.
 Similarly, the corresponding relations for the coefficients $W_{Sk}$
 introduced in \cite{RUPAK-SHORESH} are:
\be \label{eq06}
W_{Sk} = \bar{W}_k - d_k\log(\chi_S)
= W^\prime_k - d_k\log(\frac{f_0^2}{\mu^2}\chi_S) \ .
\ee
 The constants in (\ref{eq05}) and (\ref{eq06}) are given by
\be \label{eq07}
c_4 = \frac{1}{256\pi^2} \ , \hspace{1.5em}
c_5 = \frac{N_s}{256\pi^2} \ , \hspace{1.5em}
c_6 = \frac{(N_s^2+2)}{512N_s^2\pi^2} \ , \hspace{1.5em}
c_8 = \frac{(N_s^2-4)}{512N_s\pi^2} \ ,
\ee
 respectively,
\be \label{eq08}
d_4 = \frac{1}{256\pi^2} \ , \hspace{1.5em}
d_5 = \frac{N_s}{256\pi^2} \ , \hspace{1.5em}
d_6 = \frac{(N_s^2+2)}{256N_s^2\pi^2} \ , \hspace{1.5em}
d_8 = \frac{(N_s^2-4)}{256N_s\pi^2} \ .
\ee

 For the valence quark mass dependence of the (squared) pseudoscalar
 meson masses one can consider, similarly to (\ref{eq03}) and
 (\ref{eq04}), the ratios
\be \label{eq09}
Rm_{VV} \equiv \frac{m_{VV}^2}{m_{SS}^2} \ , \hspace{3em}
Rm_{VS} \equiv \frac{m_{VS}^2}{m_{SS}^2} \ .
\ee
 In the present paper we prefer to divide these ratios by the tree
 level dependences and consider
\begin{eqnarray}
Rn_{VV} \hspace*{-1.5em}&& \equiv \frac{m_{VV}^2}{\xi m_{SS}^2}
= 1-\eta_S\frac{(\xi-1)}{\xi}
\nonumber
\\[0.5em] &&
+ 8(\xi-1)\chi_S(2L_{S8}-L_{S5})
+ 8N_s\frac{(\xi-1)}{\xi}\eta_S\chi_S(L_{S4}-W_{S6})
\nonumber
\\[0.5em] &&
+ \frac{\chi_S}{16N_s\pi^2}\frac{(\xi-1)}{\xi}(\xi+\eta_S)
- \frac{\chi_S}{16N_s\pi^2}(1+2\eta_S)\log(1+\eta_S)
\nonumber
\\[0.5em] && \label{eq10}
+ \frac{\chi_S}{16N_s\pi^2}\frac{(2\xi^2-\xi-\eta_S+3\eta_S\xi)}{\xi}
\log(\xi+\eta_S) \ ,
\end{eqnarray}
 and
\begin{eqnarray}
Rn_{VS} \hspace*{-1.5em}&& \equiv \frac{2m_{VS}^2}{(\xi+1) m_{SS}^2}
= 1-\eta_S\frac{(\xi-1)}{(\xi+1)}
\nonumber
\\[0.5em] &&
+ 4(\xi-1)\chi_S(2L_{S8}-L_{S5})
+ 8N_s\frac{(\xi-1)}{(\xi+1)}\eta_S\chi_S(L_{S4}-W_{S6})
\nonumber
\\[0.5em] &&
- \frac{\chi_S}{16N_s\pi^2}(1+2\eta_S)\log(1+\eta_S)
\nonumber
\\[0.5em] && \label{eq11}
+ \frac{\chi_S}{16N_s\pi^2}\frac{(\xi^2+\xi+\eta_S+3\eta_S\xi)}{(\xi+1)}
\log(\xi+\eta_S) \ .
\end{eqnarray}

 A useful quantity is the double ratio of decay constants
 \cite{JLQCD:CHLOGF} which does not depend on any of the NLO
 coefficients.
 In other words there one can see the chiral logarithms alone.
 The NLO expansion for this quantity is:
\be \label{eq12}
RRf \equiv \frac{f_{VS}^2}{f_{VV}f_{SS}} =
1 + \frac{\chi_S}{32N_s\pi^2}(\xi-1)
- \frac{\chi_S}{32N_s\pi^2}(1+\eta_S)\log\frac{\xi+\eta_S}{1+\eta_S} \ .
\ee

 The double ratio of the pion mass squares \cite{JLQCD:CHLOGM}
 corresponding to (\ref{eq10}) and (\ref{eq11}) has the NLO
 expansion
\begin{eqnarray}
RRn \hspace*{-1.5em}&& \equiv 
\frac{4\xi m_{VS}^4}{(\xi+1)^2m_{VV}^2 m_{SS}^2}
= 1 - \frac{\eta_S(\xi-1)^2}{\xi(\xi+1)}
\nonumber
\\[0.5em] &&
+ \frac{\chi_S(\xi^2+\xi+\eta_S+3\eta_S\xi^2)\log(\xi+\eta_S)}
{16N_s\pi^2\xi(\xi+1)}
- \frac{\chi_S(2\eta_S+1)\log(1+\eta_S)}{16N_s\pi^2}
\nonumber
\\[0.5em] && \label{eq13}
- \frac{\chi_S(\xi-1)(\xi+\eta_S)}{16N_s\pi^2\xi}
+ \frac{8N_s\chi_S\eta_S(\xi-1)^2}{\xi(\xi+1)}(L_{S4}-W_{S6}) \ .
\end{eqnarray}
%

\subsection{Quadratic corrections}\label{sec2.1}

 A complete NNLO (i.~e.~two-loop) calculation in PQChPT for our
 physical quantities is not yet available.
 Nevertheless, the general form of NNLO tree-graph (``counter\-term
 insertion'') contributions can be given \cite{SHARPE-VDWATER}\footnote{
 We thank the authors for communicating us the content of this paper
 prior to publication.}.
 For instance, one has for the pion mass square:
\be \label{eq14}
\frac{\delta m_{AB}^2}{m_{AB}^2} \;=\; 
\alpha_1 \chi_S^2 + \alpha_2 \chi_S(\chi_A+\chi_B) +
\alpha_3 (\chi_A+\chi_B)^2 + \alpha_4 (\chi_A^2+\chi_B^2) \ .
\ee
 (Here A and B denote generic quark indices: S is the label for the sea
 quarks, V for valence quarks.)
 For the pion decay constant there is a similar expression. 
 This information is very useful in order to estimate the importance of
 the NNLO terms in our present range of quark masses.

 The general characteristics of the NNLO terms is that they are
 proportional to the quark mass square: $\chi_S^2$.
 (Here we only consider terms in the continuum limit and hence neglect
 lattice artifacts.
 This will be to some extent justified a posteriori by the observed
 smallness of ${\cal O}(a)$ terms.)
 Neglecting loop contributions, which are at the NLO order relatively
 small, the dependence on the quark mass ratio $\xi$ is at most
 quadratic and can, therefore, be represented by terms proportional to
 $(\xi-1)$ and $(\xi-1)^2$.
 Therefore, these contributions have the generic form
\be \label{eq15}
D_X \chi_S^2 (\xi-1) + Q_X \chi_S^2 (\xi-1)^2 \ .
\ee
 Here $X$ denotes an index specifying the considered ratio as, for
 instance, $X=fVV,\; nVS$ etc. for the single ratios and
 $X=fd$ and $X=nd$ for the double ratios $RRf$ and $RRn$, respectively.
 The NLO tree-graph contributions for the single ratios $Rf$ and $Rn$
 are also proportional to $(\xi-1)$.
 These can be parametrized as $L_X \chi_S (\xi-1)$ (for instance, we
 have $L_{fVV} \equiv 4L_5$ and $L_{nVV} \equiv 8(2L_8-L_5)$).
 The inclusion of $D_X$-type terms is equivalent to a
 linear dependence of the effective $L_X$'s for fixed $\chi_S$:
\be \label{eq16}
L_X^{eff} = L_X+D_X \chi_S \ .
\ee
 At this point one has to remember that mathematically speaking --
 in order to completely remove the effect of higher order terms --
 $L_X$ is defined in the limit $\chi_S \to 0$.

 The NNLO coefficients are not all independent but satisfy the
 relations
\begin{eqnarray}
D_{fVS} = \half D_{fVV} \ ,  & &
D_{nVS} = \half D_{nVV} \ ,
\nonumber 
\\[1em]
 D_{fd} = 0 \ , & & D_{nd} = 0 \ ,
\nonumber 
\\[1em] \label{eq17}
Q_{fd} = 2Q_{fVS}-Q_{fVV}+\frac{1}{4}L_{fVV}^2 \ , & &
Q_{nd} = 2Q_{nVS}-Q_{nVV}+\frac{1}{4}L_{nVV}^2 \ .
\end{eqnarray}
 The first line is a consequence of the general structure of the NNLO
 tree-graph contributions.
 The last two lines follow from the definition of $RRf$ and $RRn$ if
 one only considers NLO and NNLO tree-graph contributions.

 We shall see in section \ref{sec4} that in our range of quark masses
 the NNLO tree-graph contributions of the form (\ref{eq15}) are
 important but can be approximately determined by global fits.
 In this way the NLO constants $L_k$ are better determined.
 Observe that a determination of the $D_X$'s is only possible in our
 analysis if different sea quark masses are included (see below).

\subsection{${\cal O}(a^2)$ corrections}\label{sec2.2}

 The idea of including leading lattice artifacts in the low energy
 effective Lagrangian for the Wilson lattice action can be extended
 to higher orders in lattice spacing.
 Indeed, in writing this paper we have seen two recent publications
 about the inclusion of ${\cal O}(a^2)$ corrections \cite{BARUSH,AOKI}.
 The general formulas derived in these papers for the ${\cal O}(a^2)$
 terms imply that in the formulas for the pion mass-squared ratios
 (\ref{eq10}), (\ref{eq11}) and (\ref{eq13}) there are only very little
 changes.
 In fact, the changes can be summarized by the replacement
\be \label{eq18}
\eta_S(L_{S4}-W_{S6}) \;\longrightarrow\; \eta_S(L_{S4}-W_{S6}) +
\frac{\eta_S^2}{N_s}(N_s W_{S4} + W_{S5} - 2 N_s W^\prime_{S6} -
2W^\prime_{S8}) \ .
\ee
 Here $W_{S6}^\prime$ and $W_{S8}^\prime$ denote some new low energy
 constants appearing in the ${\cal O}(a^2)$ part of the effective
 Lagrangian.
 This means that fitting the valence quark mass dependence with our
 formulas (\ref{eq10}), (\ref{eq11}) and (\ref{eq13}) effectively takes
 into account also ${\cal O}(a^2)$ corrections.

 Concerning the ratios of the pion decay constants in (\ref{eq03}),
 (\ref{eq04}) and (\ref{eq12}) the situation is expected to be similar
 but there, in addition to the ${\cal O}(a^2)$ terms, also new types of
 ${\cal O}(am_q)$ terms may appear.
 
\section{Sea quark mass dependence}\label{sec3}

 The dependence on the sea quark mass can be treated similarly to the
 valence quark mass dependence considered in section \ref{sec2}.
 Here one chooses a ``reference value'' of the sea quark mass
 $\chi_R$ and determines the ratios of the coupling and decay constant
 as a function of
\be \label{eq19}
\sigma   \equiv \frac{\chi_S}{\chi_R} \ , \hspace{3em}
\tau \equiv \frac{\rho_S}{\rho_R}  \ .
\ee
 Instead of $\tau$ one can also use
\be \label{eq20}
\eta_S \equiv \frac{\rho_S}{\chi_S} \ , \hspace{3em}
\eta_R \equiv \frac{\rho_R}{\chi_R}
\ee
 which satisfy
\be \label{eq21}
\frac{\tau}{\sigma} = \frac{\eta_S}{\eta_R} \ .
\ee
 With this we have for the decay constants
\begin{eqnarray}
Rf_{SS} \hspace*{-1.5em}&& \equiv \frac{f_{SS}}{f_{RR}}
= 1 + 4(\sigma-1)\chi_R (N_s L_{R4}+L_{R5})
+ 4(\eta_S\sigma-\eta_R)\chi_R (N_s W_{R4}+W_{R5})
\nonumber
\\[0.5em] && \label{eq22}
-\frac{N_s\chi_R}{32\pi^2}\sigma(1+\eta_S)\log[\sigma(1+\eta_S)]
+\frac{N_s\chi_R}{32\pi^2}(1+\eta_R)\log(1+\eta_R)
\end{eqnarray}
 and for the mass squares
\begin{eqnarray}
Rn_{SS} \hspace*{-1.5em}&& \equiv \frac{m_{SS}^2}{\sigma m_{RR}^2}
= 1 + \eta_S - \eta_R
+8(\sigma-1)\chi_R(2N_s L_{R6}+2L_{R8}-N_s L_{R4}-L_{R5})
\nonumber
\\[0.5em] &&
+8(\eta_S\sigma-\eta_R)\chi_R
(2N_s W_{R6}+2W_{R8}-N_s W_{R4}-W_{R5}-N_s L_{R4}-L_{R5})
\nonumber
\\[0.5em] && \label{eq23}
+\frac{\chi_R}{16\pi^2 N_s}\sigma(1+2\eta_S)\log[\sigma(1+\eta_S)]
-\frac{\chi_R}{16\pi^2 N_s}(1+2\eta_R)\log(1+\eta_R) \ .
\end{eqnarray}
 Of course, the coefficients $L_{Rk}$ and $W_{Rk}\; (k=4,5,6,8)$
 are now defined at the scale $f_0\sqrt{\chi_R}$ therefore in the
 relations (\ref{eq05}) and (\ref{eq06}) $\chi_S$ is replaced by
 $\chi_R$.

 The logarithmic dependence of $L_{Sk}$'s and $W_{Sk}$'s have to be
 taken into account also in simultaneous fits of the valence quark mass
 dependence at several sea quark mass values.
 Choosing a fixed {\em reference sea quark mass} $\chi_R$ we have from
 (\ref{eq05}) and (\ref{eq06}) with $\mu=f_0\sqrt{\chi_R}$
\be \label{eq24}
L_{Sk} = L_{Rk}-c_k\log\sigma \ ,  \hspace{3em}
W_{Sk} = W_{Rk}-d_k\log\sigma \ .
\ee
 The NLO PQChPT formulas for the valence quark mass dependence
 in terms of the reference sea quark mass are obtained by the following
 subsitutions in (\ref{eq03}), (\ref{eq04}), (\ref{eq10})-(\ref{eq13}):
\begin{eqnarray}
\chi_S \to \sigma\chi_R \ ,\hspace{3em}  L_{Sk} &\to& L_{Rk} \ ,
\hspace{3em}
W_{Sk} \to  W_{Rk} \ ,
\nonumber 
\\
\log(1+\eta_S)      &\to& \log[\sigma(1+\eta_S)] \ ,
\nonumber 
\\
\log(\xi+\eta_S)    &\to& \log[\sigma(\xi+\eta_S)] \ ,
\nonumber 
\\ \label{eq25}
\log(1+\xi+2\eta_S) &\to& \log[\sigma(1+\xi+2\eta_S)] \ .
\end{eqnarray}

 An important feature of both the valence- and sea-quark mass
 dependences considered in the present work is that they are ratios
 taken at a fixed value of the gauge coupling ($\beta$).
 These are renormalization group invariants independent from the
 $Z$-factors of multiplicative renormalization since the $Z$-factors
 only depend on the gauge coupling and not on the quark mass.
 Taking ratios of pion mass squares and pion decay constants at varying
 quark masses has, in general, the advantage that quark mass independent
 corrections -- for instance of ${\cal O}(a)$ and/or ${\cal O}(a^2)$ --
 cancel.

\section{Numerical simulations}\label{sec4}

 We performed Monte Carlo simulations with $N_s=2$ degenerate sea quarks
 on a $16^4$ lattice at $\beta=5.1$ and three values of $\kappa$:
 $\kappa_0=0.176$, $\kappa_1=0.1765$ and $\kappa_2=0.177$.
 For the {\em reference sea quark mass} we choose
 $\kappa_R \equiv \kappa_0=0.176$.
 A summary of the simulation points is reported in table~\ref{tab01},
 where also the set-up of the TSMB algorithm for the different
 simulation points can be found.
 The gauge field configurations collected for the evaluation of the
 physical quantities are separated by 10 TSMB update cycles consisting
 out of boson field and gauge field updates and noisy correction steps.
 It turned out that these configurations were statistically independent
 from the point of view of almost all secondary quantities considered.
 Exceptions are $r_0/a$ and $M_r$ (see below) where autocorrelation
 lengths of 2-5 units in the configuration sequences appear.
\begin{table}
\begin{center}
\parbox{12cm}{\caption{\label{tab01}\em
 Parameters of the simulations: all simulations were done at
 $\beta=5.10$ with determinant breakup $N_f=1+1$.
 The other TSMB-parameters are: the interval of polynomial
 approximations $[\epsilon,\lambda]$ and the polynomial orders
 $n_{1,2,3}$ \rm\protect\cite{TSMB}.}}
\end{center}
\begin{center}
\begin{tabular}{|l|c|c||c|c|c|c|c|}\hline
run&$\kappa$&configurations &$\epsilon$&$\lambda$&$n_1$&$n_2$&$n_3$
\\\hline\hline
0 & $0.1760$ & $ 1811$ & $4.50\cdot10^{-4}$ & $3.0$ & $40$ &
 $210$ & $220$  
\\\hline
1 & $0.1765$ & $  746$ & $2.50\cdot10^{-4}$ & $3.0$ & $40-44$ &
 $280$ & $260-340$ 
\\\hline
2 & $0.1770$ & $ 1031$ & $3.75\cdot10^{-5}$ & $3.0$ & $54$ &
 $690$ & $840$ 
\\\hline
\end{tabular}
\end{center}
\end{table}

 We investigated for each simulation point the valence quark mass 
 dependence of the pseudo-Goldostone boson spectrum and decay constants; 
 the values of the valence $\kappa$ considered for each simulation point
 are reported in table~\ref{tab02}.
 In these intervals the valence quark masses are approximately changing
 in the range $\half m_{sea} \leq m_{valence} \leq 2 m_{sea}$.

 A rough estimate of the sea quark mass range can be obtained by
 considering the quantity $M_r\equiv (r_0 m_\pi)^2$, which for the
 strange quark gives $M_r \approx 3.1$.
 (Here $r_0 \approx 0.5\, {\rm fm}$ is the Sommer scale parameter which
 characterizes the distance scale intrinsic to the gauge field.)
 In our simulation points the value of $M_r$ ranges between
 $M_r\approx 2.10$ and $M_r\approx 1.09$, corresponding to about
 $\frac{2}{3}$ and  $\frac{1}{3}$ of the value for the strange quark
 mass.
 Since the valence quark masses roughly go down to
 $m_{valence} \simeq \half m_{sea}$, they reach
 $m_{valence} \simeq \frac{1}{6} m_s$.
 In our configuration samples we did not encounter problems with
 ``exceptional gauge configurations'' -- in spite of the smallness of
 the valence quark mass.
 This means that the quark determinant effectively suppresses such
 configurations.

 Standard methods for the extraction of the relevant physical quantities
 have been applied (a more detailed description is given in our previous
 paper \cite{NF2TEST} and in \cite{GEBERT}).
 Statistical errors have been obtained by the {\em linearization method}
 \cite{ALPHA:BENCHMARK,WOLFF} which we found more reliable than
 jack-knifing on bin averages.
\begin{table}
\begin{center}
\parbox{12cm}{\caption{\label{tab02}\em
 Values of the valence quark hopping parameter.}}
\end{center}
\begin{center}
\begin{tabular}{|l|c|c|c|}\hline
  run                                     
& 0        & 1        & 2       
\\\hline\hline
  $\kappa_{\mbox{{\scriptsize sea}}}$     
& $0.1760$ & $0.1765$ & $0.1770$
\\\hline
  $\kappa_{\mbox{{\scriptsize valence}}}$ 
& $0.1685$ & $0.1710$ & $0.1743$ \\
& $0.1705$ & $0.1718$ & $0.1747$ \\
& $0.1720$ & $0.1726$ & $0.1751$ \\
& $0.1730$ & $0.1734$ & $0.1754$ \\
& $0.1735$ & $0.1742$ & $0.1759$ \\
& $0.1745$ & $0.1750$ & $0.1763$ \\
& $0.1750$ & $0.1758$ & $0.1767$ \\
& $0.1770$ & $0.1772$ & $0.1775$ \\
& $0.1775$ & $0.1778$ & $0.1779$ \\
& $0.1785$ & $0.1785$ & $0.1783$ \\
& $0.1790$ & $0.1791$ & $0.1787$ \\
& $0.1800$ & $0.1797$ & $0.1791$ 
\\\hline
\end{tabular}
\end{center}
\vspace*{-1em}
\end{table}

 Within a mass-independent renormalization scheme - defined at zero
 quark mass - the $Z$-factors of multiplicative renormalization depend
 only on the gauge coupling ($\beta$) and not on the quark mass
 ($\kappa$).
 Similarly, the lattice spacing $a$ is also a function of the gauge
 coupling alone \cite{BANGALORE}.
 Therefore, since our simulation points are at fixed gauge coupling
 $\beta=5.1$, the ratios of the sea quark masses can be obtained
 by taking ratios of the measured bare quark masses in lattice units
 $Z_q am_q$.
 Here $Z_q$ is the multiplicative renormalization factor for the
 quark mass which is the ratio of the Z-factors of the pseudoscalar
 density and axialvector current ($Z_q = Z_P/Z_A$) because we determine
 the quark mass by the PCAC-relation: $m_q \equiv m_q^{PCAC}$
 \cite{NF2TEST}.
 (Of course, in the valence quark ratios the factor $Z_q a$ also cancels
 trivially.)
 The obtained values of the sea quark mass ratios
 $\sigma_i \equiv m_{qi}/m_{q0}$ ($i=1,2$) are given in
 table~\ref{tab03} together with some other basic quantities.

 Note that by identifying the quark mass ratios in the ChPT formulas
 with the ratios of the PCAC quark masses (``axialvector Ward identity
 quark masses'') one assumes that these two kinds of renormalized quark
 masses are proportional to each other.
 As it is shown, for instance, by eq.~(48) in \cite{AOKI} this is indeed
 the case -- apart from lattice artifacts of ${\cal O}(am_q)$ and
 ${\cal O}(a^2)$.
 The quark mass independent part of the ${\cal O}(a^2)$ terms are
 cancelled by taking ratios.
 The remaining quark mass dependent lattice artifacts are neglected in
 the present paper.
 
 The critical value of the hopping parameter where the quark mass
 vanishes can be estimated by a quadratic extrapolation using the values
 of $\sigma_{1,2}$:
\be \label{eq26}
\sigma_i \equiv \frac{m_{qi}}{m_{q0}} =
\frac{(\kappa_i^{-1}-\kappa_{cr}^{-1}) + 
d_\sigma(\kappa_i^{-1}-\kappa_{cr}^{-1})^2}
{(\kappa_0^{-1}-\kappa_{cr}^{-1}) + 
d_\sigma(\kappa_0^{-1}-\kappa_{cr}^{-1})^2} \ .
\ee
 The values of $\sigma_{1,2}$ in table~\ref{tab03} give the solution:
 $\kappa_{cr}=0.1773(1)$ and $d_\sigma = -11.2(8)$.
 (The relatively large absolute value of $d_\sigma$ shows that the
 quadratic term in the extrapolation is important.)

 The value of the lattice spacing $a$ can be inferred from the value
 of $r_0/a$ at $\kappa=\kappa_{cr}$.
 This can also be determined by a quadratic extrapolation of the
 values of $r_0/a$ given in table~\ref{tab03} with the result:
 $r_0(\kappa_{cr})/a = 2.65(7)$.
 Taking, by definition, $r_0(\kappa_{cr})=0.5\, {\rm fm}$ this gives
 for the lattice spacing: $a=0.189(5) \, {\rm fm}$.

 The physical volume following from the lattice spacing is comfortably
 large: $L \simeq 3.0\, {\rm fm}$.
 Since the minimal value of the pion mass in lattice units in our points
 is $am_\pi^{min} \simeq 0.43$ for sea quarks and
 $am_\pi^{min} \simeq 0.30$ for the lightest valence quark, we have
 $Lm_\pi \geq 4.8$.

 Another information given by the values of $M_r$ is an estimate of the
 quark mass parameter $\chi_S$ in the ChPT formulas.
 For instance, in the reference point we have from $r_0f_0 \simeq 0.23$
 \cite{DURR}: $\chi_R^{estimate} \approx M_r/(r_0 f_0)^2 \simeq 39.8$.
\begin{table}[ht]
\begin{center}
\parbox{12cm}{\caption{\label{tab03}\em
 The values of some basic quantities in our simulation points.
 Statistical errors in last digits are given in parentheses.}}
\end{center}
\begin{center}
\begin{tabular}[ht]{|c||c|c|c|}
\hline
$\kappa$                 & $\kappa_0$  & $\kappa_1$  & $\kappa_2$
\\ \hline\hline
$r_0/a$                  & 2.149(15)   & 2.171(88)   & 2.395(52)
\\ \hline
$am_\pi$                 & 0.6747(14)  & 0.6211(22)  & 0.4354(68)
\\ \hline
$M_r=(r_0 m_\pi)^2$      & 2.103(26)   & 1.824(41)   & 1.088(47)
\\ \hline
$Z_q am_q$               & 0.07472(32) & 0.06247(51) & 0.03087(36)
\\ \hline
$\sigma_i=m_{qi}/m_{q0}$ & 1.0         & 0.8361(52)  & 0.4132(34)
\\ \hline
\end{tabular}
\end{center}
\end{table}

\subsection{Valence quark mass dependence}\label{sec4.1}

 For a fixed value of the sea quark mass $\chi_S$ the valence quark
 mass dependence of the ratios $Rf_{VV,VS}$, $Rn_{VV,VS}$, $RRf$ and
 $RRn$ is determined by five parameters:
\be \label{eq27}
\chi_S \ ,  \hspace{1em}
\chi_S L_{S5} \ ,  \hspace{1em}
\chi_S L_{S85} \equiv \chi_S (2L_{S8}-L_{S5}) \ ,  \hspace{1em}
\chi_S L_{S4W6} \equiv \chi_S (L_{S4}-W_{S6})\ ,  \hspace{1em}
\eta_S \ .
\ee
 The dependence is linear in the first four of them but it is non-linear
 in $\eta_S$.

 After performing such fits of the data we realized that the sea quark
 mass dependence is not consistent with the NLO PQChPT formulas.
 In particular, the best fit values of the $\chi_S$'s have ratios
 considerably closer to 1 than $\sigma_{1,2}$ in table~\ref{tab03}
 and the change of the $L_k$'s with $\chi_S$ is also not consistent
 with (\ref{eq24}).
 This shows that NNLO effects are important and, therefore, we tried
 fits including NNLO tree-graph terms of the form given in (\ref{eq15}).
 The list of the relevant NNLO parameters is:
\be \label{eq28}
\chi_R^2 D_{fVV,nVV} \ ,  \hspace{3em}
\chi_R^2 Q_{fVV,fVS,fd,nVV,nVS,nd} \ .
\ee
 $Q_{fd}$ and $Q_{nd}$ have to satisfy the quadratic relations given in
 the last line of (\ref{eq17}) but in order to keep linearity we did not
 impose these relations and fitted the eight parameters in (\ref{eq28})
 independently.
 After performing the fits one can check how well the relations for
 $Q_{fd}$ and $Q_{nd}$ are fulfilled. 

 The global fit of the valence quark mass dependence for several
 values of the sea quark mass has twelve linear parameters: the first
 four in (\ref{eq27}) with $\chi_S$ replaced by $\chi_R$
\be \label{eq29}
\chi_R \ ,  \hspace{1em}
\chi_R L_{R5} \ ,  \hspace{1em}
\chi_R L_{R85} \equiv \chi_R (2L_{R8}-L_{R5}) \ ,  \hspace{1em}
\chi_R L_{R4W6} \equiv \chi_R (L_{R4}-W_{R6})
\ee
 and the eight in (\ref{eq28}).
 In addition there are the non-linear parameters, in our case three
 of them: $\eta_S=\eta_{0,1,2}$.
\begin{table}[ht]
\begin{center}
\parbox{12cm}{\caption{\label{tab04}\em
 Values of best fit parameters for the valence quark mass dependence.
 Quantities directly used in the fitting procedure are in bold face.}}
\end{center}
\begin{center}
\begin{tabular}{|c|r||c|r|}\hline
\bteq{\chi_R}           & $33.5(2.4)$            &           &
\\\hline
\bteq{\chi_RL_{R4W6}}   & $5.24(38)\cdot10^{-2}$ & $L_{R4W6}$&
 $1.564(71)\cdot10^{-3}$
\\\hline
\bteq{\chi_R^2Q_{nd}}   & $6.5(1.8)\cdot10^{-3}$ & $Q_{nd}$  &
 $5.80(79)\cdot10^{-6}$   
\\\hline
\bteq{\chi_RL_{R5}}     & $10.06(44)\cdot10^{-2}$ & $L_{R5}$  &
 $3.00(19)\cdot10^{-3}$   
\\\hline
\bteq{\chi_R^2D_{fVV}}  & $-9.3(1.7)\cdot10^{-2}$& $D_{fVV}$ &
 $-8.3(1.9)\cdot10^{-5}$  
\\\hline
\bteq{\chi_R^2Q_{fVV}}  & $-2.80(19)\cdot10^{-2}$& $Q_{fVV}$ &
 $-2.50(50)\cdot10^{-5}$  
\\\hline
\bteq{\chi_R^2Q_{fVS}}  & $-2.197(45)\cdot10^{-2}$& $Q_{fVS}$ &
 $-1.96(29)\cdot10^{-5}$  
\\\hline
\bteq{\chi_R^2Q_{fd}}   & $-0.99(14)\cdot10^{-2}$& $Q_{fd}$  &
 $-0.89(45)\cdot10^{-5}$ 
\\\hline
\bteq{\chi_RL_{R85}}    & $-2.10(12)\cdot10^{-2}$& $L_{R85}$ &
 $-6.25(52)\cdot10^{-4}$  
\\\hline
\bteq{\chi_R^2D_{nVV}}  & $-1.67(20)\cdot10^{-1}$& $D_{nVV}$ &
 $-1.49(10)\cdot10^{-4}$  
\\\hline
\bteq{\chi_R^2Q_{nVV}}  & $-8.44(67)\cdot10^{-2}$& $Q_{nVV}$ &
 $-7.53(48)\cdot10^{-5}$  
\\\hline
\bteq{\chi_R^2Q_{nVS}}  & $-4.05(25)\cdot10^{-2}$& $Q_{nVS}$ &
 $-3.61(29)\cdot10^{-5}$  
\\\hline
\end{tabular}
\end{center}
\end{table}

 Multi-parameter linear fits are easy and, except for degenerate
 situations, the chi-square always has a unique well-defined minimum.
 Non-linear fits involving the $\eta$'s are more problematic,
 therefore we adopted the following procedure: performing non-linear
 fits at individual sea quark mass values we obtained the starting
 values of $\eta_{0,1,2}$.
 Then for fixed values of $\eta_{0,1,2}$ we performed a linear fit of
 the twelve parameters in (\ref{eq28})-(\ref{eq29}) and looked for a
 minimum of the chi-square as a function of $\eta_{0,1,2}$.
 For the sea quark masses we imposed the relation
 $\chi_S=\sigma\chi_R$ and for the NLO parameters the relations in
 (\ref{eq24}) with the values of $\sigma_{1,2}$ given in
 table~\ref{tab03}.
 (The possible dependence of the NNLO parameters $D$ and $Q$ on
 $\sigma$ has been neglected.)
 The minimum of the chi-square after the non-linear minimization is
 near
\be \label{eq30}
\eta_0 = 0.07 \ ,  \hspace{2em}
\eta_1 = 0.03 \ ,  \hspace{2em}
\eta_2 = 0.02 \ .
\ee
 The minimum as a function of $\eta_{0,1,2}$ is rather shallow but
 definitely within the bounds $0 \leq \eta_{0,1,2} \leq 0.10$.
 The minimization of the chi-square of the linear fit does not
 change the $\eta$'s substantially: already the starting values
 are close to (\ref{eq30}).
 This confirms the small value of $\eta_S$ found in our previous paper
 at $\beta=4.68$ \cite{VALENCE}.

 In contrast to the stable values of the $\eta$'s there are large
 fluctuations in the basic parameter $\chi_R$: one can obtain values
 in the range $13 \leq \chi_R \leq 40$ depending on the set of functions
 fitted, on the fit interval etc.
 This is presumably the effect of our small number (only three) of sea
 quark masses.
 In order to obtain more stable results we fixed $\eta_{0,1,2}$
 according to (\ref{eq30}) and first determined in a linear fit
 the three parameters $\chi_R,\; \chi_R L_{R4W6}$ and $\chi_R^2 Q_{nd}$
 from $RRn$.
 These parameters were then used as an input in the linear fit of
 the remaining nine parameters.

\begin{figure}
\begin{center}
\includegraphics[width=4.5cm]{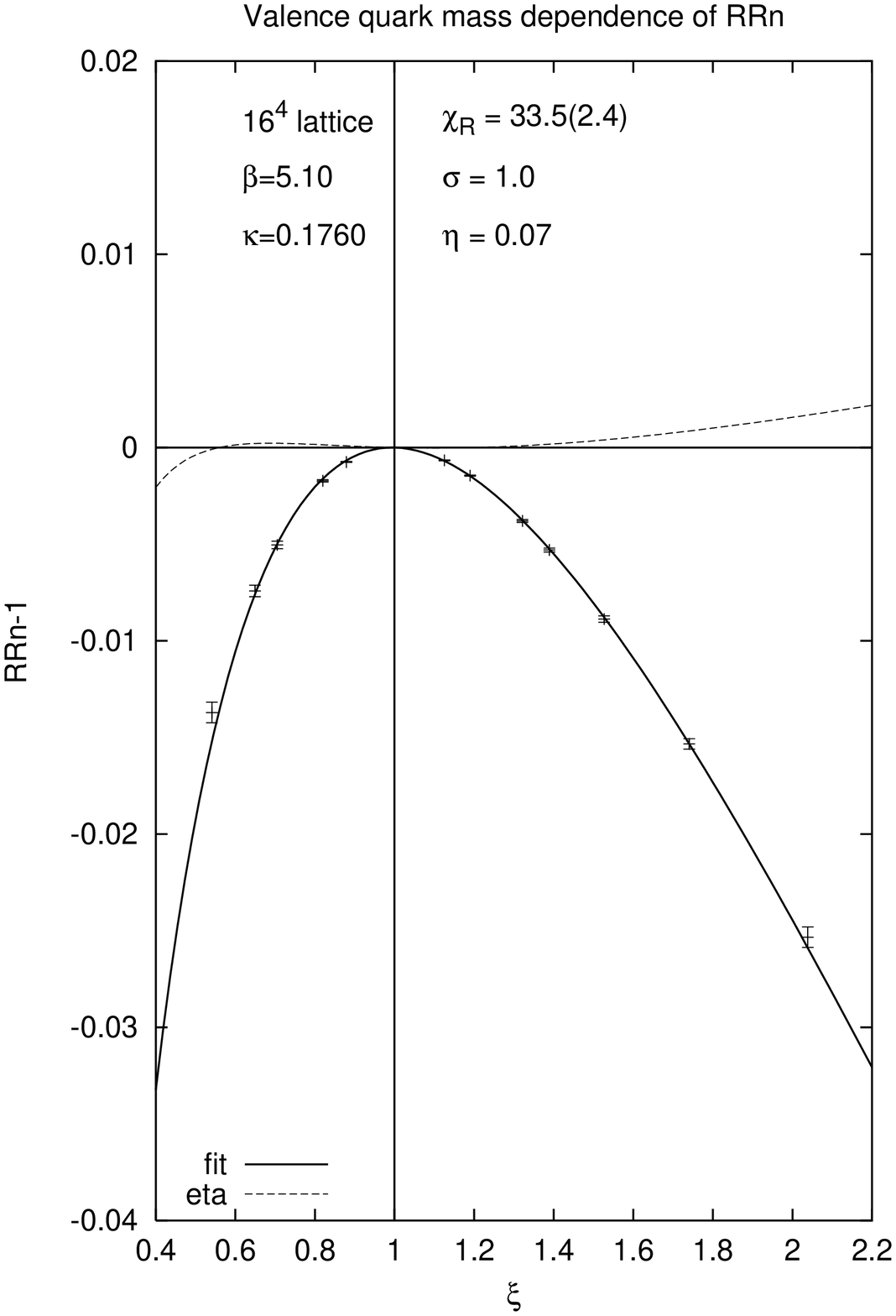}
\includegraphics[width=4.5cm]{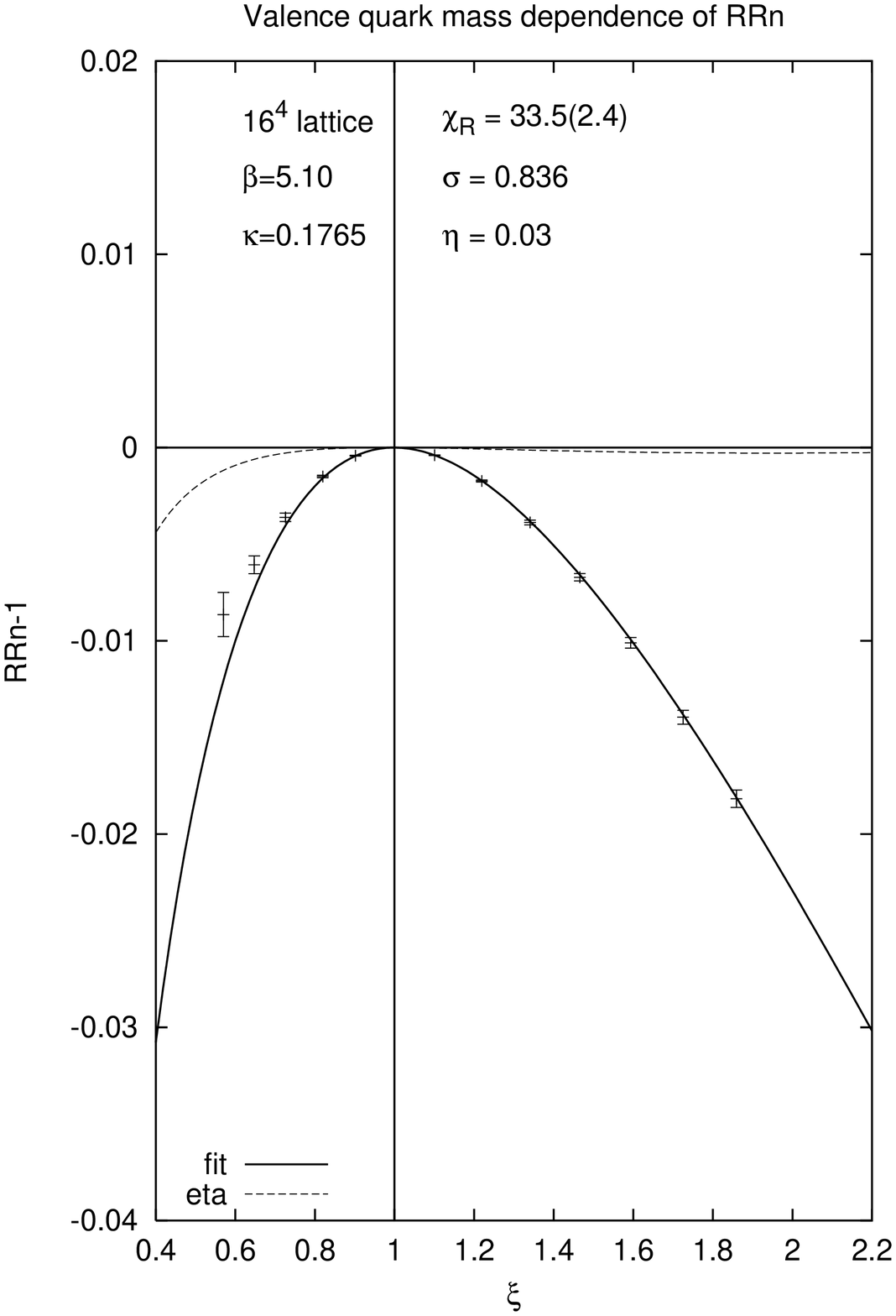}
\includegraphics[width=4.5cm]{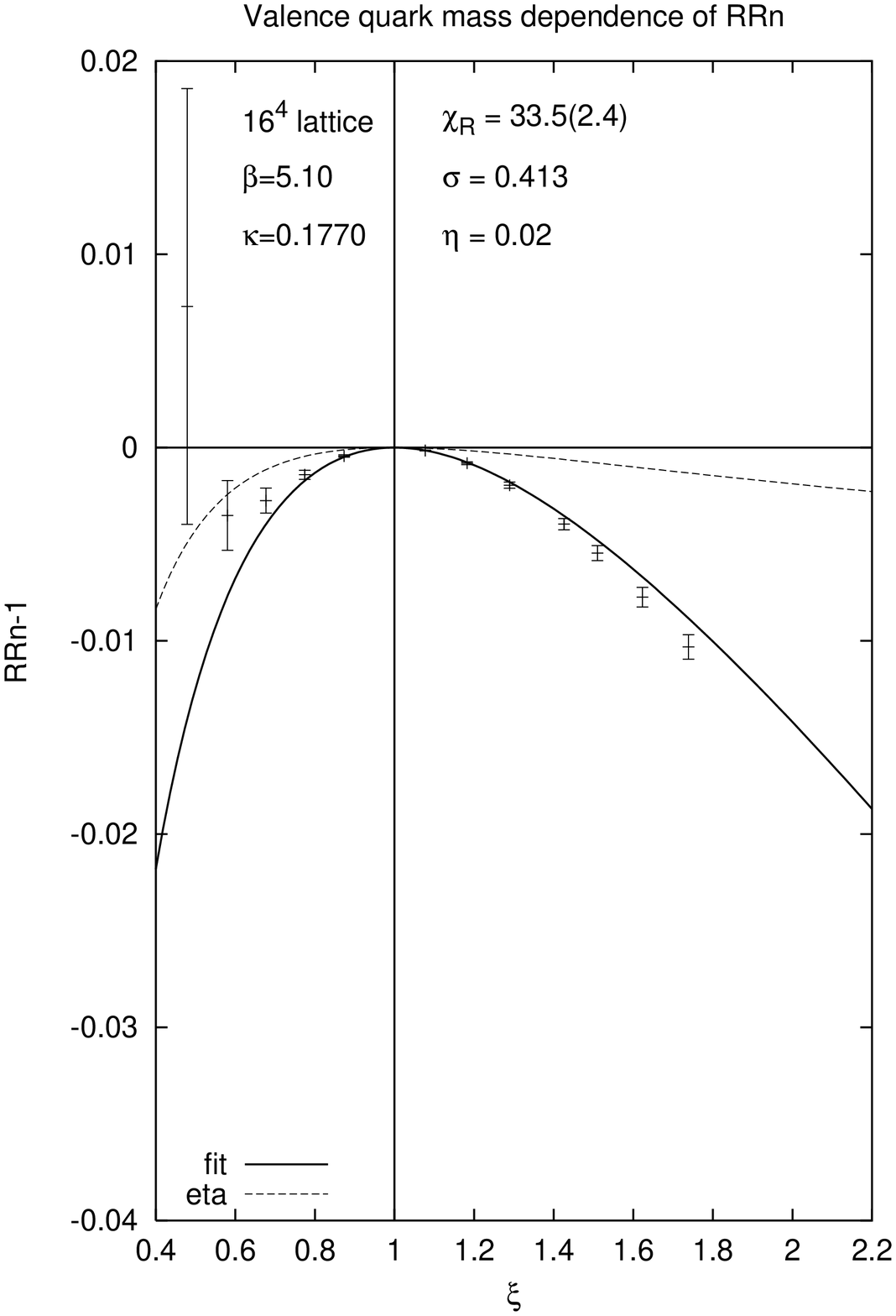}
\includegraphics[width=4.5cm]{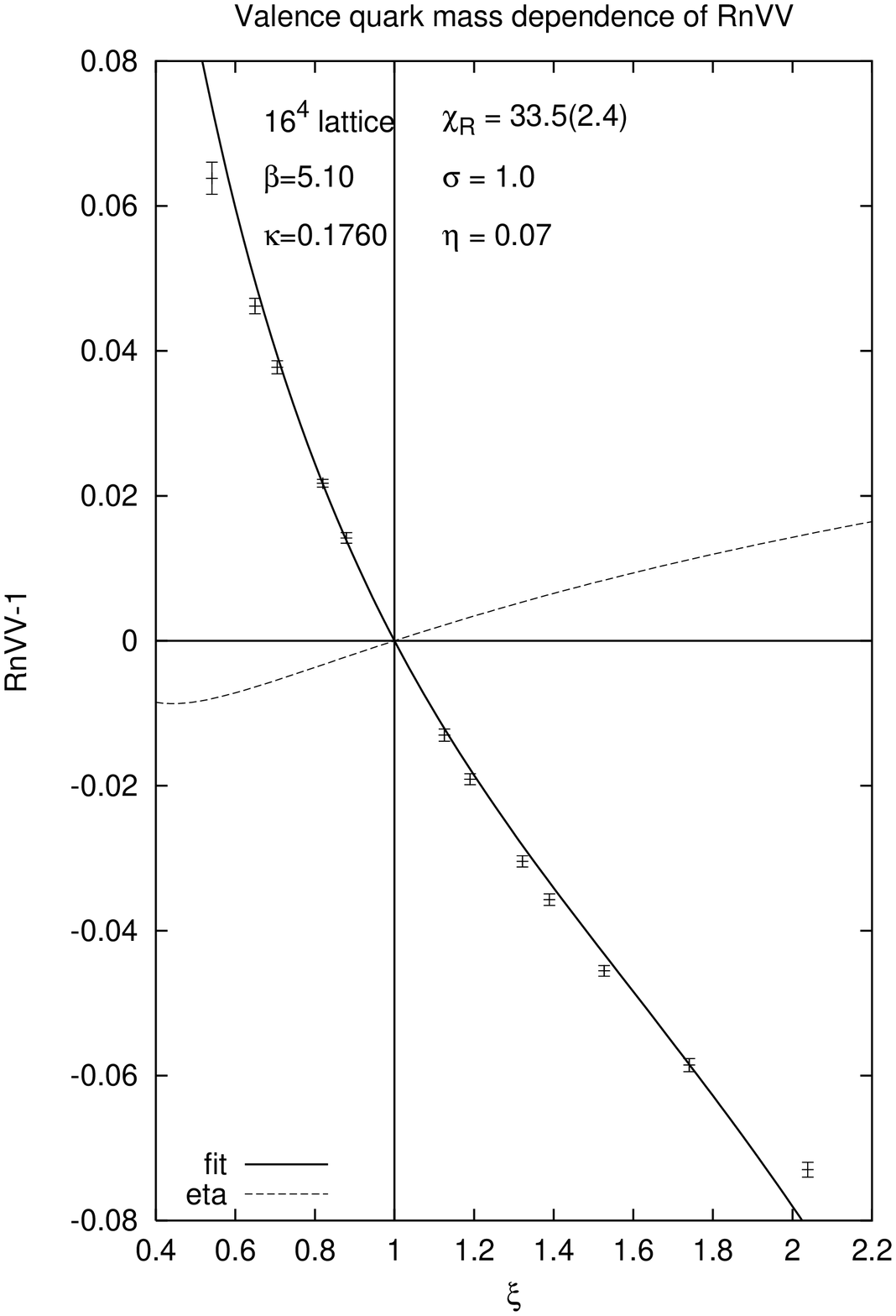}
\includegraphics[width=4.5cm]{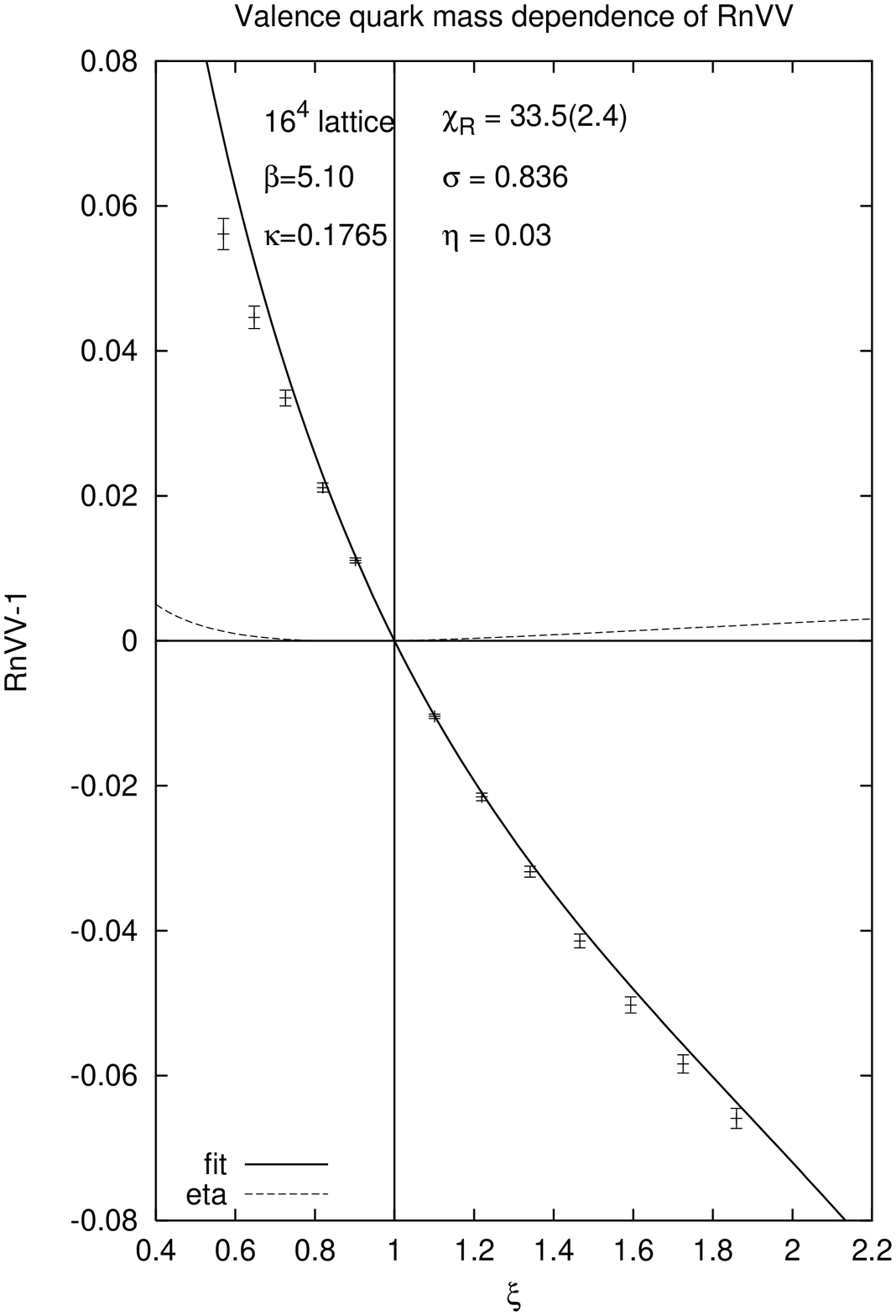}
\includegraphics[width=4.5cm]{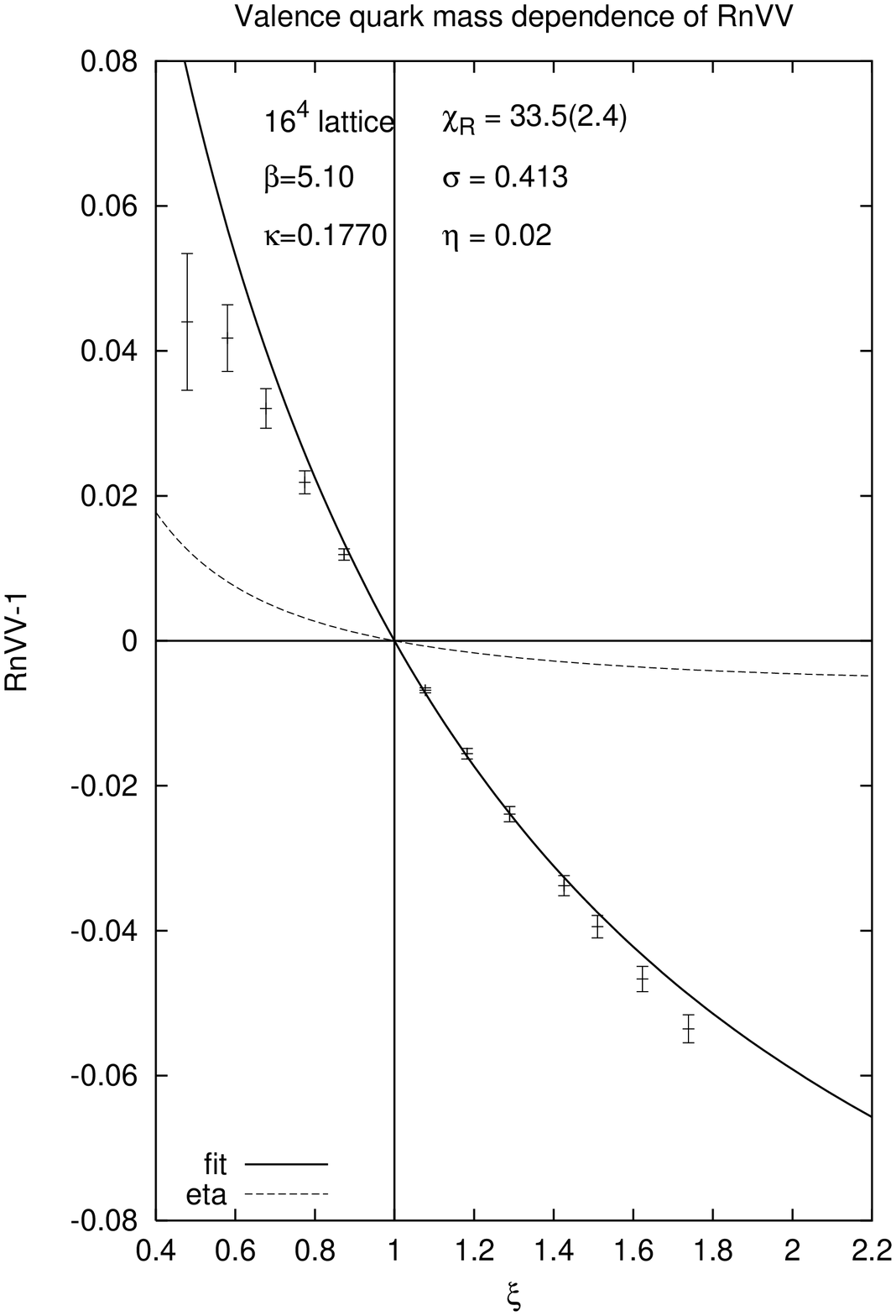}
\includegraphics[width=4.5cm]{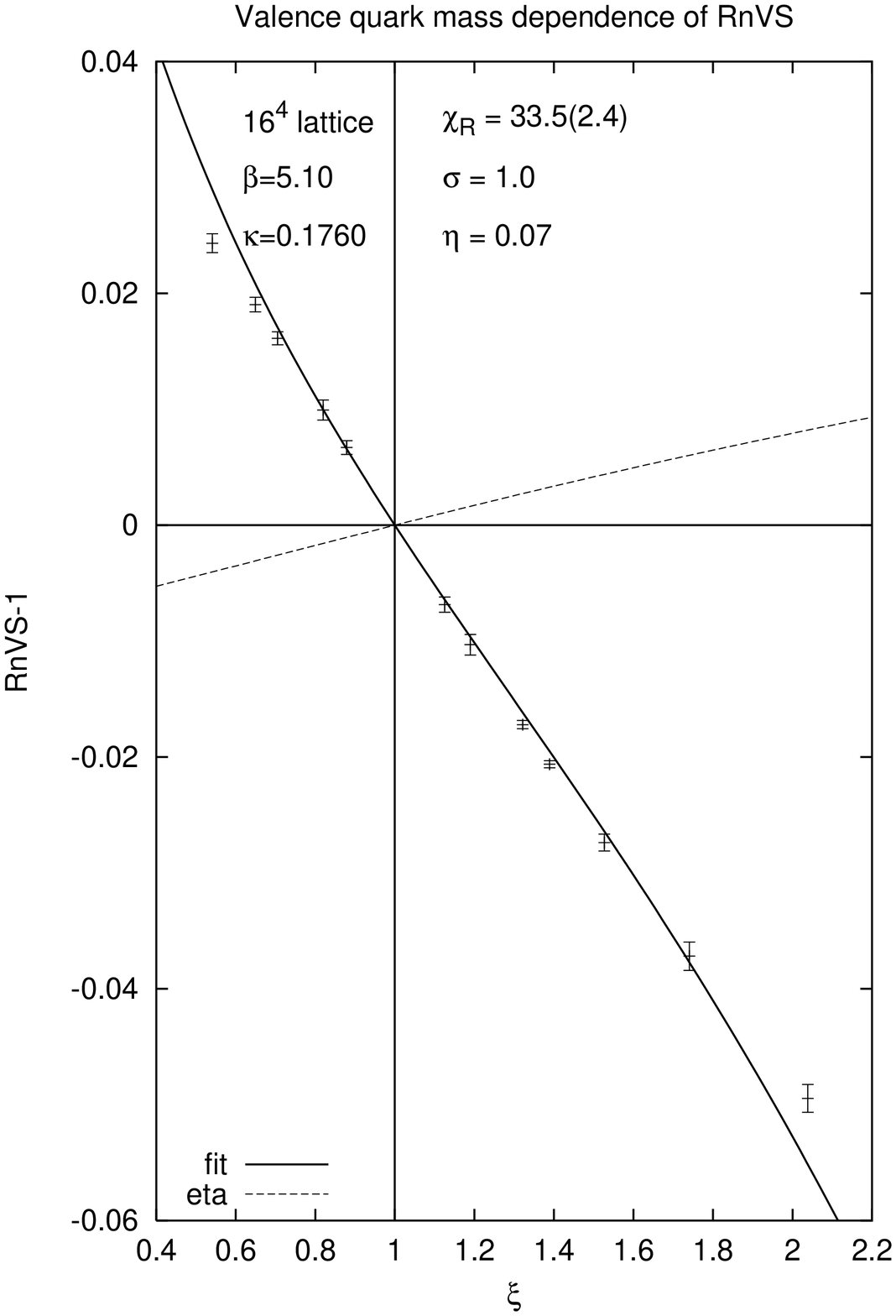}
\includegraphics[width=4.5cm]{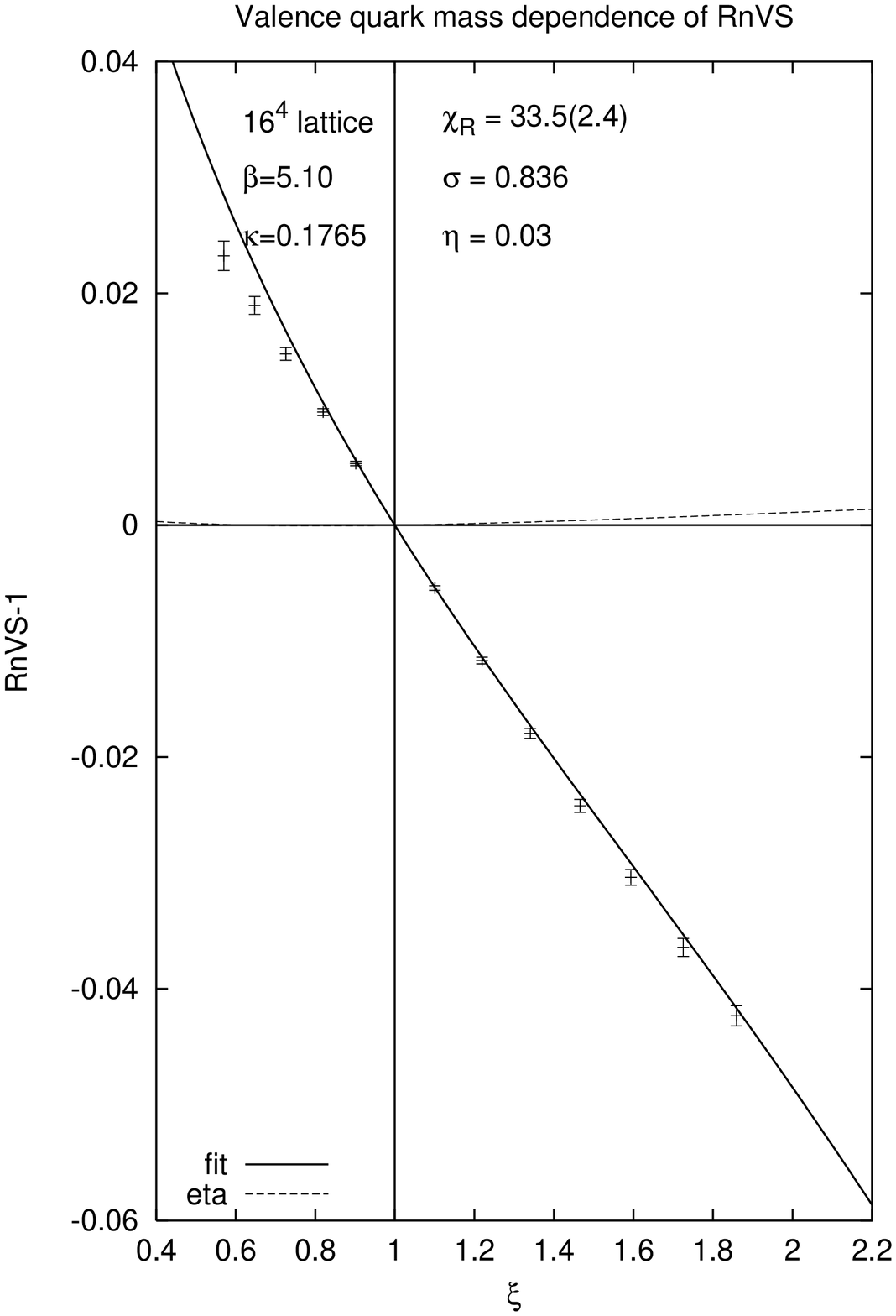}
\includegraphics[width=4.5cm]{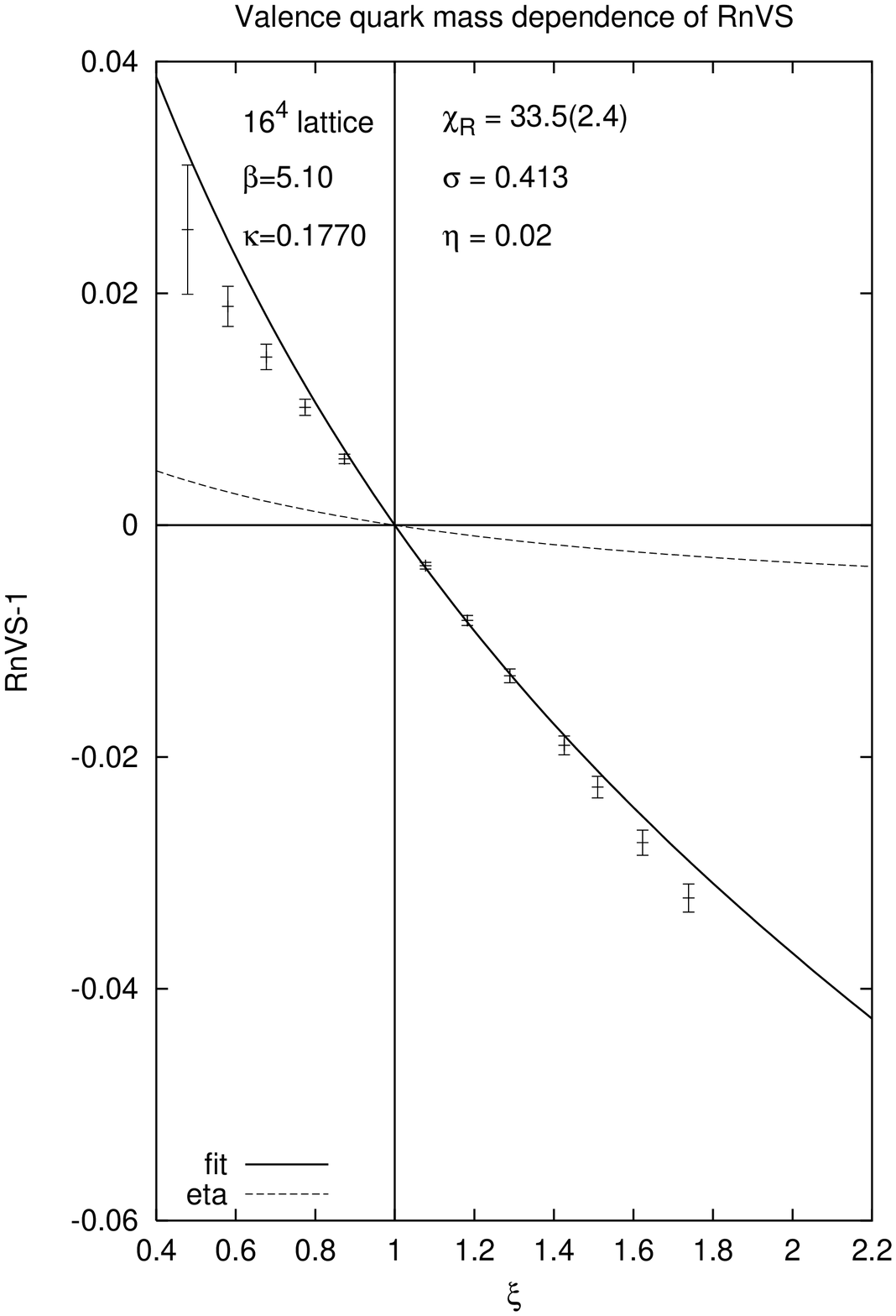}
\end{center}
\begin{center}
\parbox{12cm}{\caption{\label{fig01}\em
 $(RRn-1)$, $(Rn_{VV}-1)$ and $(Rn_{VS}-1)$ for the three different
 sea quark mass values (full lines).
 Beside the fit the unphysical contribution (proportional to $\eta_S$) is
 separately shown (broken lines).}}
\end{center}
\end{figure}

\begin{figure}
\begin{center}
\includegraphics[width=4.5cm]{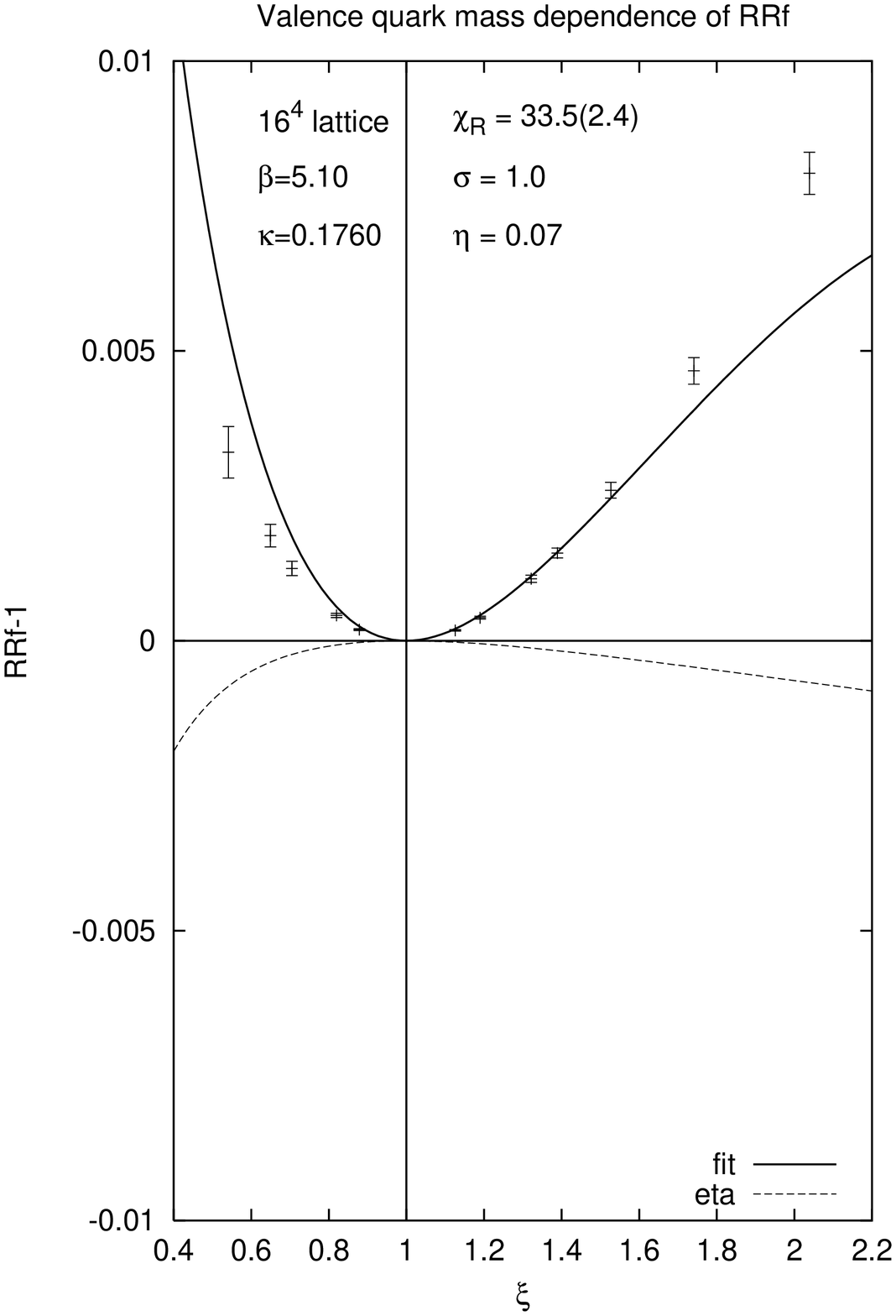}
\includegraphics[width=4.5cm]{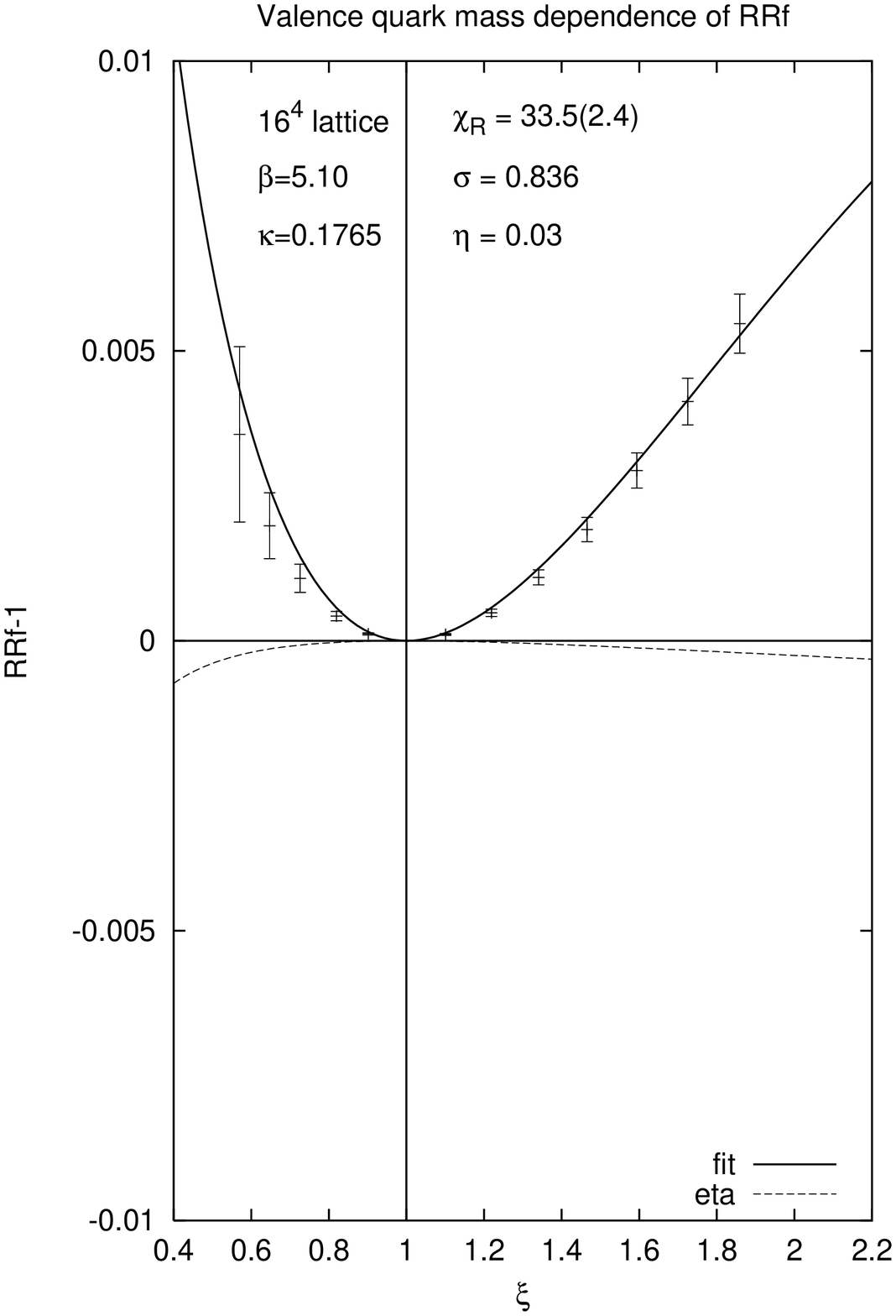}
\includegraphics[width=4.5cm]{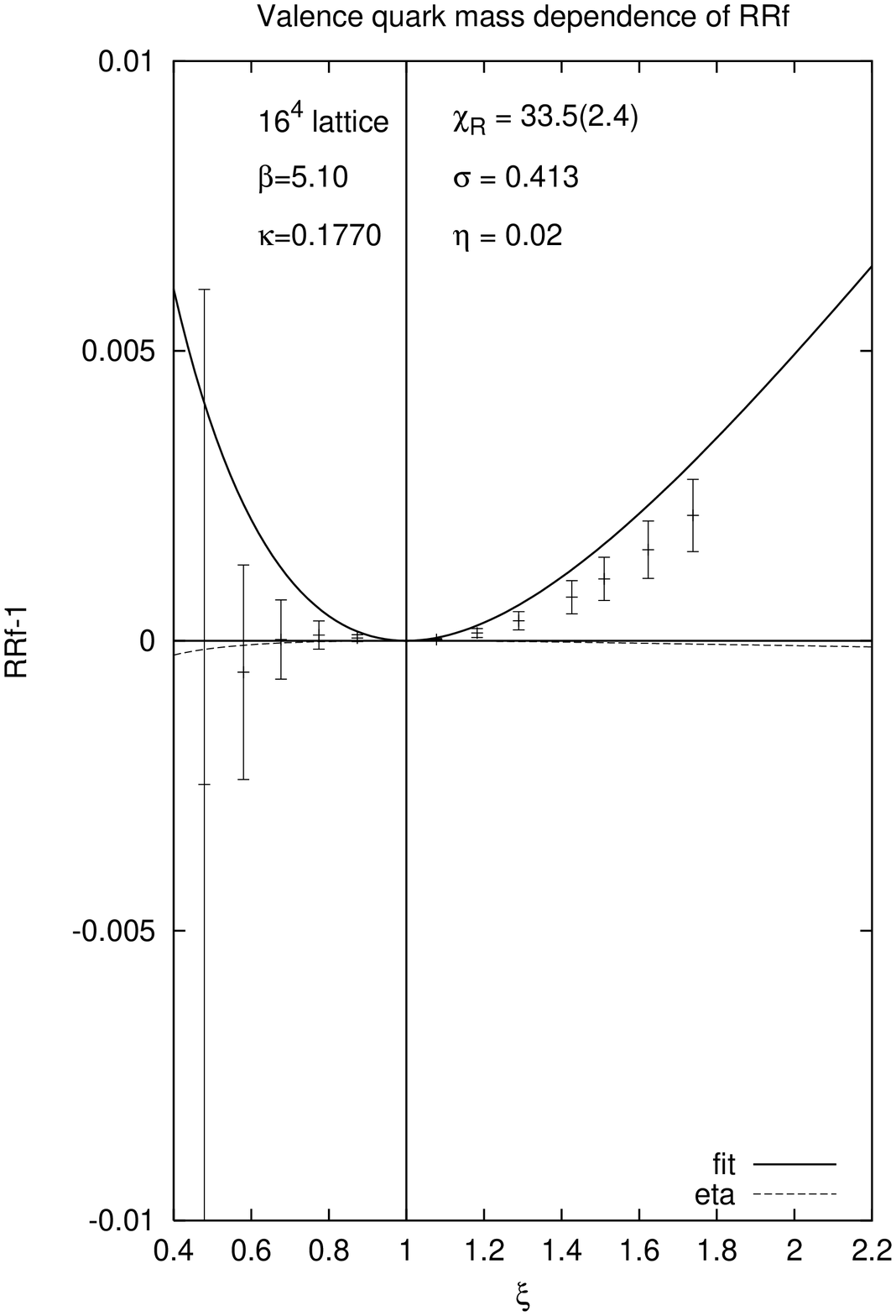}
\includegraphics[width=4.5cm]{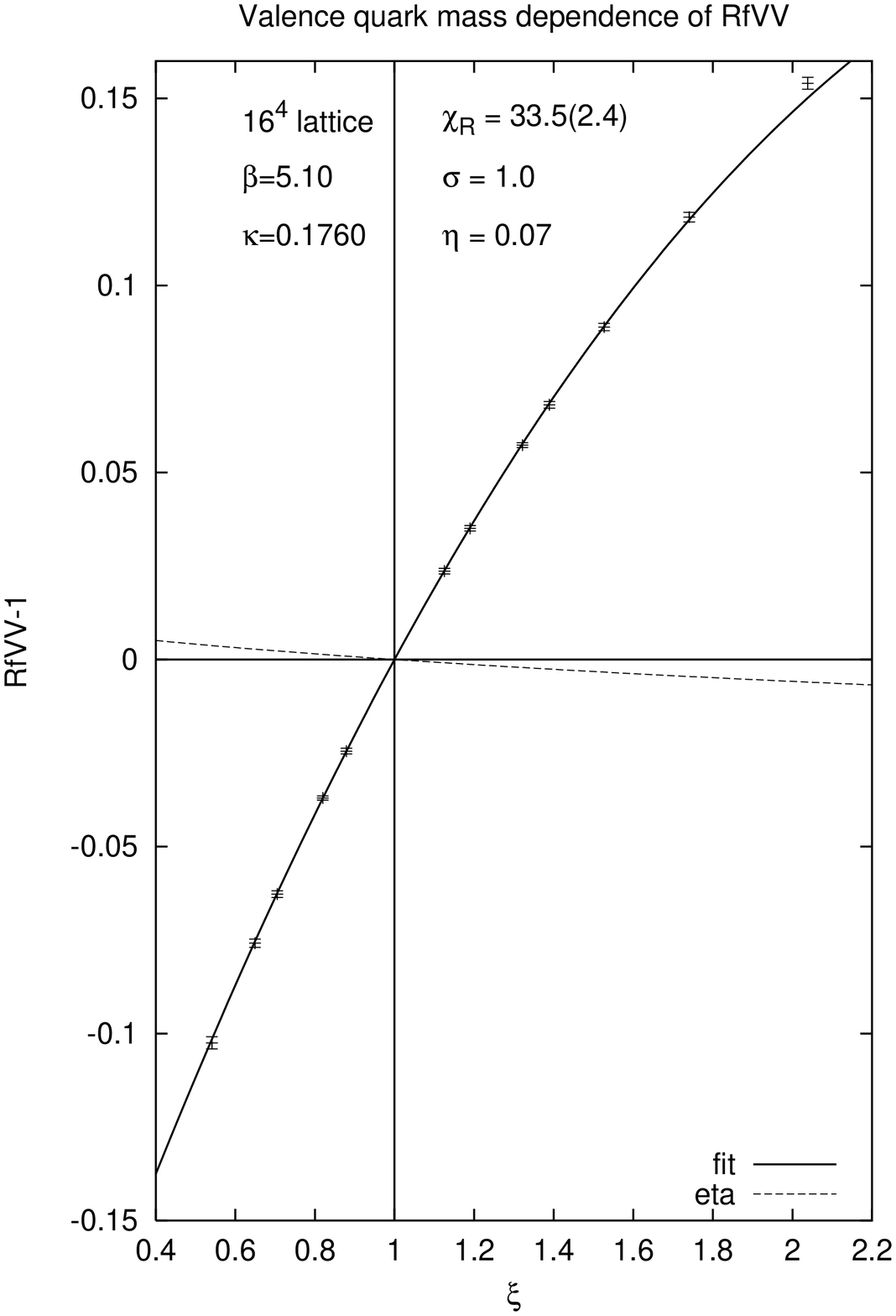}
\includegraphics[width=4.5cm]{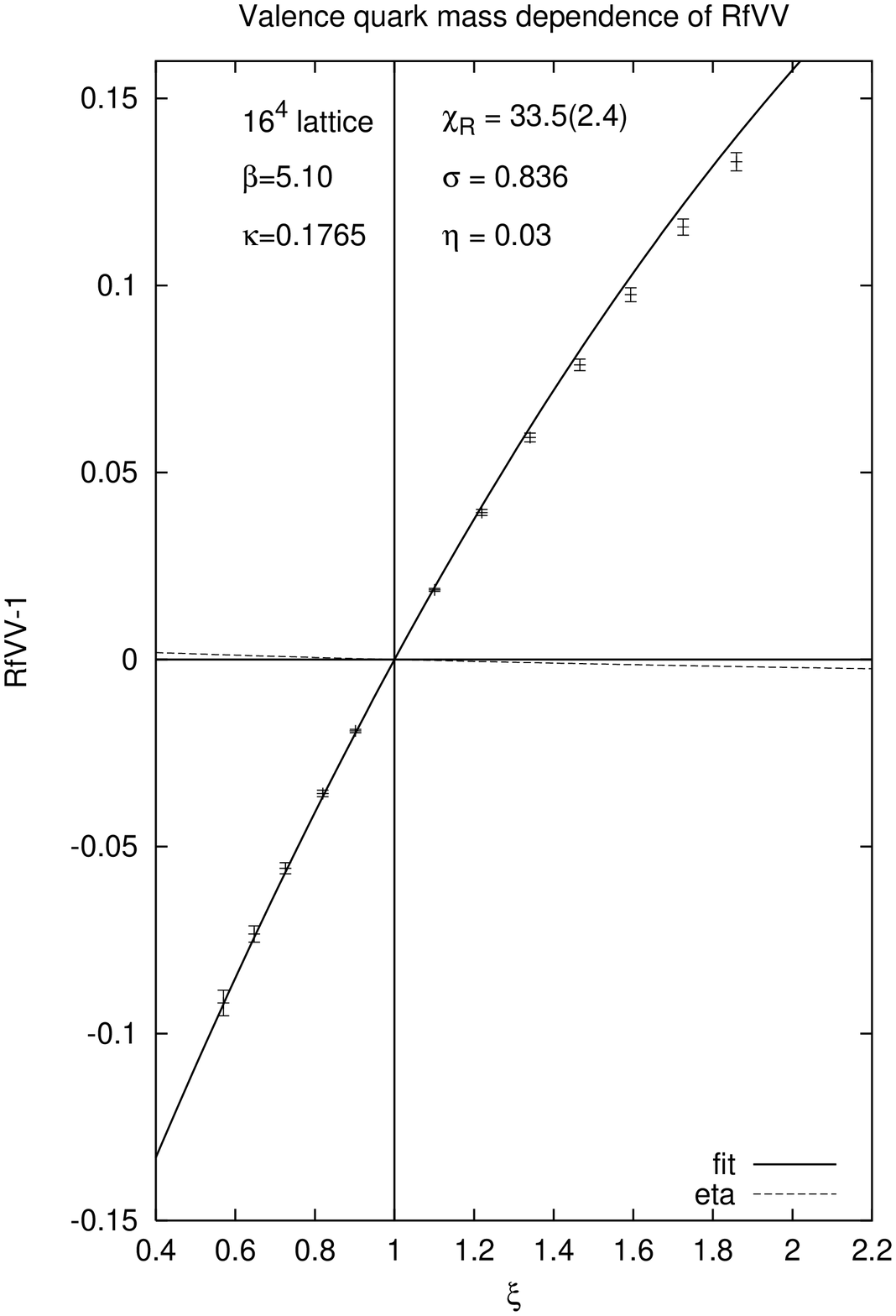}
\includegraphics[width=4.5cm]{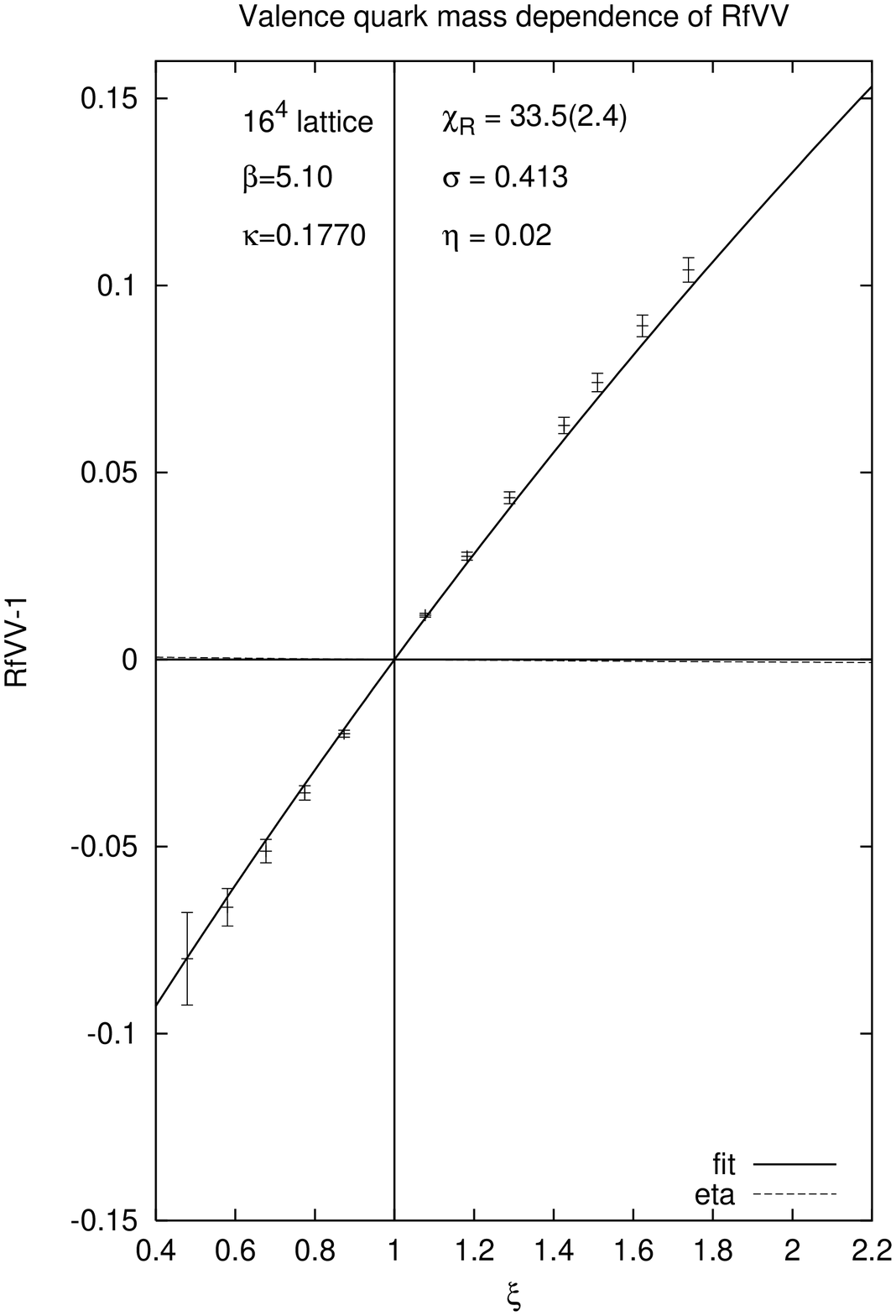}
\includegraphics[width=4.5cm]{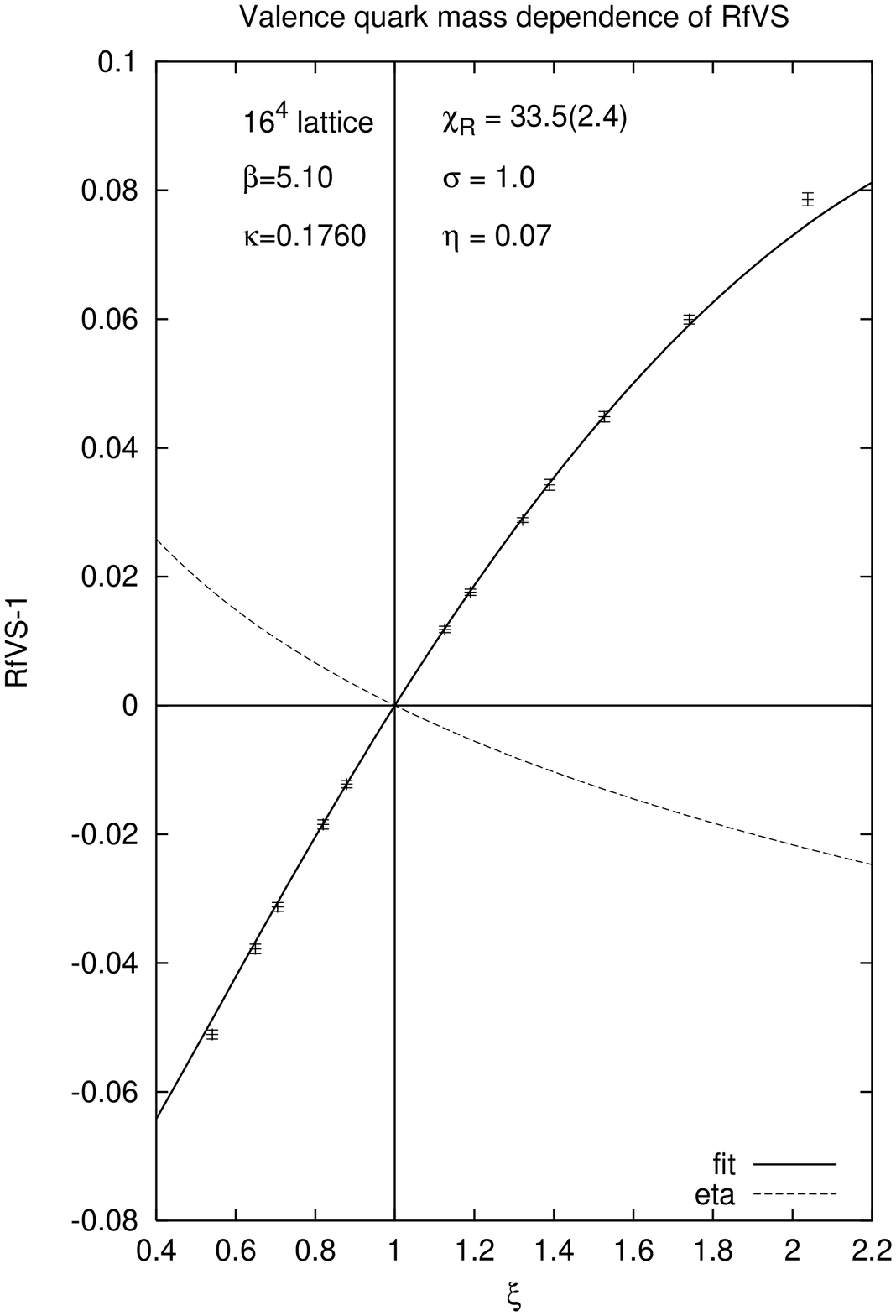}
\includegraphics[width=4.5cm]{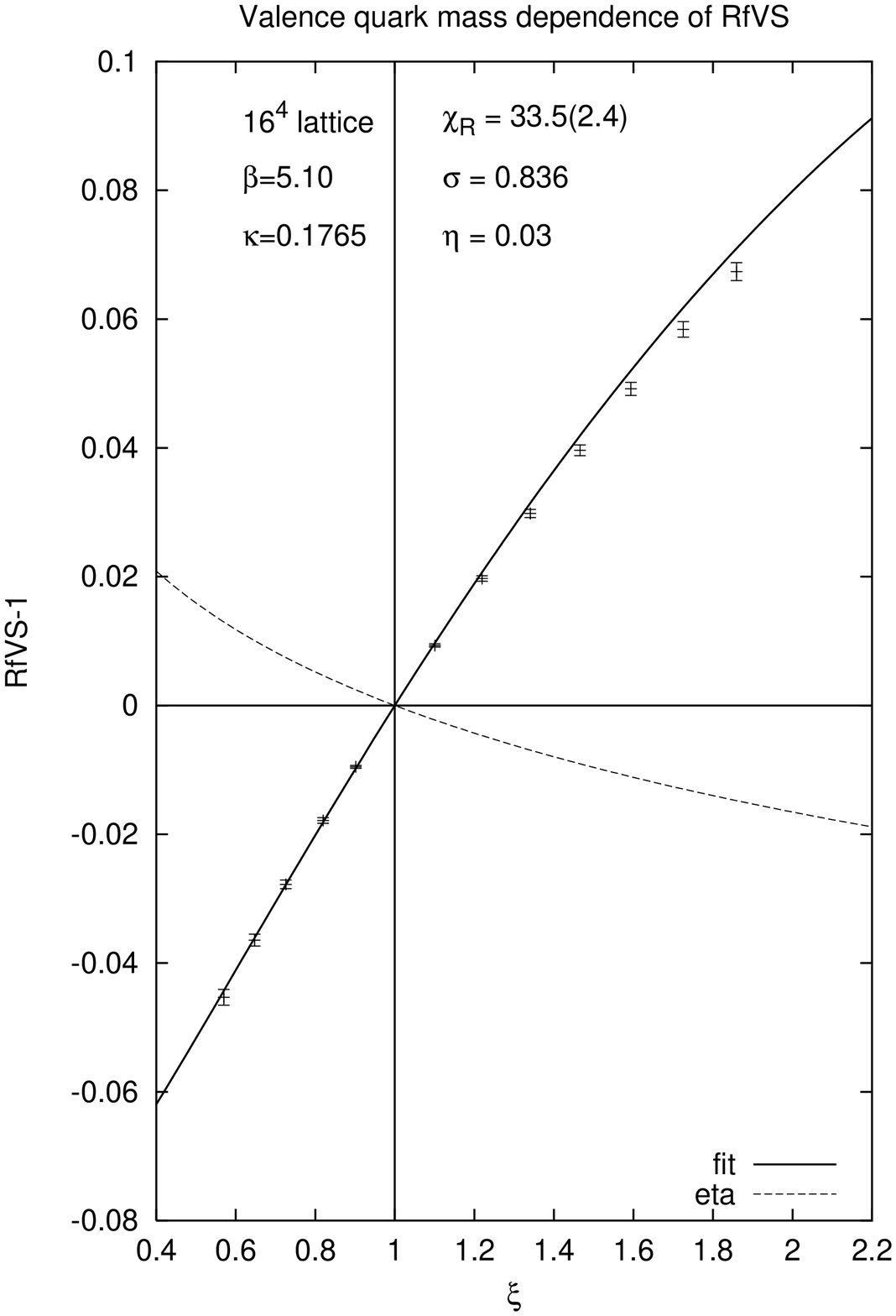}
\includegraphics[width=4.5cm]{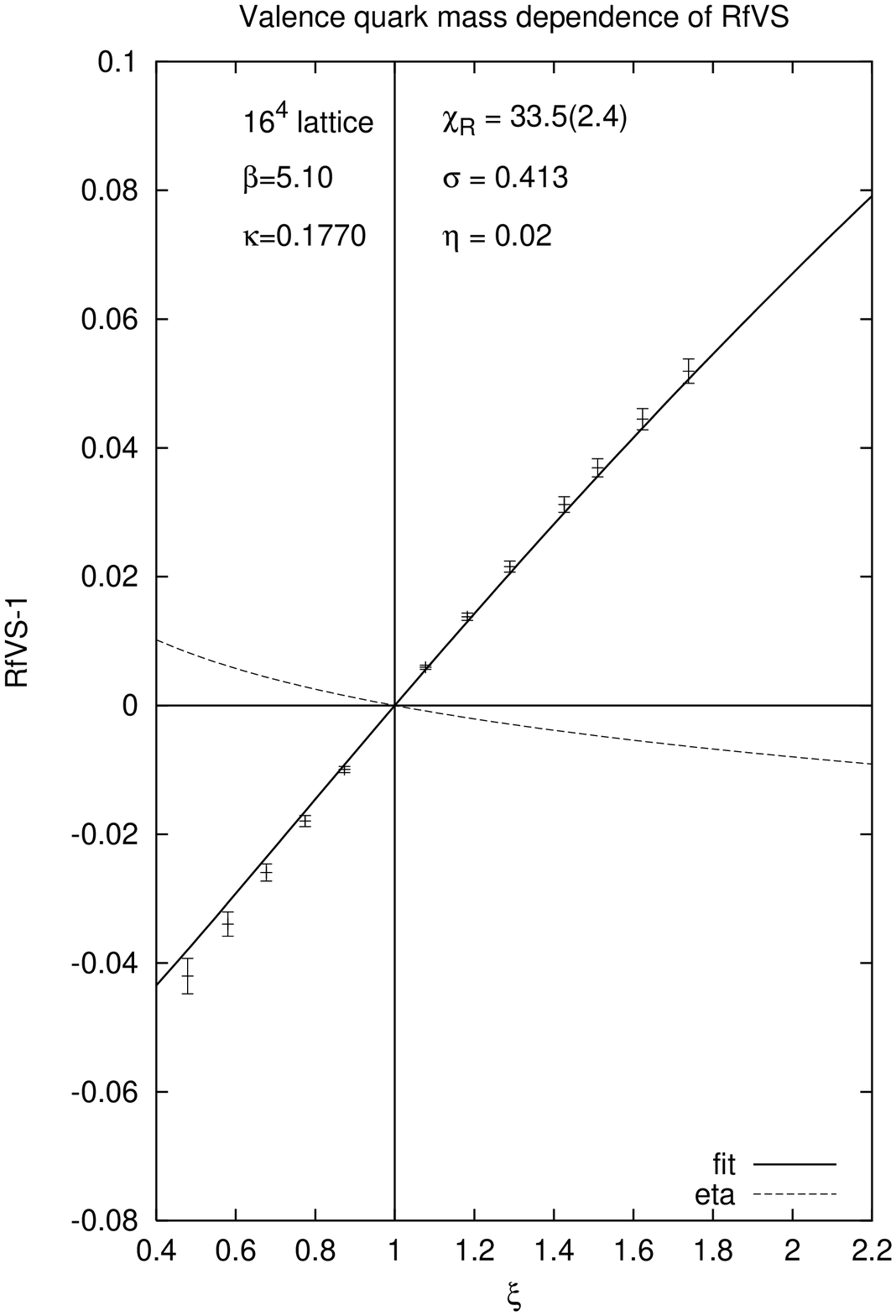}
\end{center}
\begin{center}
\parbox{12cm}{\caption{\label{fig02}\em
 $(RRf-1)$, $(Rf_{VV}-1)$ and $(Rf_{VS}-1)$ for the three different
 sea quark mass values (full lines).
 Beside the fit the unphysical contribution (proportional to $\eta_S$) is
 separately shown (broken lines).}}
\end{center}
\end{figure}

 All 18 valence quark mass dependences considered can be reasonably well
 fitted.
 The best fit is shown by figures~\ref{fig01} and \ref{fig02}.
 The sum of the chi-squares of the linear fits is $\chi^2 \simeq 300$
 for a number of degrees of freedom $n.d.f.=18 \cdot 12-12 = 204$.
 Most of the chi-squares comes from the points with largest
 and smallest valence quark masses where there are obviously some
 systematic deviations, too.
 The parameters of best fit are given in table~\ref{tab04}.
 The values in the table show that there are some discrepancies in both
 relations in the last line of (\ref{eq17}) but the deviations are
 not very large.
 The first and second relation give:
 $-0.89(49) \cdot 10^{-5} \simeq 2.05(39) \cdot 10^{-5}$ and
 $0.52(9) \cdot 10^{-5} \simeq 0.92(8) \cdot 10^{-5}$, respectively.
\begin{table}[ht]
\begin{center}
\parbox{16cm}{\caption{\label{tab05}\em
 Values of combinations of $\alpha_k$'s obtained from the best fit
 values in table~\protect\ref{tab04} and \protect\ref{tab06}.}}
\end{center}
\begin{center}
\begin{tabular}{|c|r||c|r||c|r|}\hline
$\alpha_5$                                                   &
 $2.24(20)$ &                 &             &                       &
\\\hline
$\alpha_{85}\equiv(2\alpha_8-\alpha_5)$                      &
 $0.762(49)$ &                &             & $(\alpha_4-\omega_6)$ &
 $2.36(9)\;\,$
\\\hline
$\alpha_{45}\equiv(2\alpha_4-\alpha_5)$                      &
 $2.40(26)$ & $\Lambda_4/f_0$ & $22.9(1.5)$ & $\omega_{45}$         &
 $-1.7(1.8)$
\\\hline
$\alpha_{6845}\equiv(4\alpha_6+2\alpha_8-2\alpha_4-\alpha_5)$&
 $0.658(86)$ & $\Lambda_3/f_0$ & $6.51(57)$  & $\omega_{6845}$       &
 $-5.43(60)$
\\\hline
\end{tabular}
\end{center}
\end{table}

 The values of the NLO and NNLO parameters themselves are also shown
 in the right hand part of table~\ref{tab04}, with errors determined
 (as always) by the linearization method.
 With the help of the formulas in (\ref{eq05})-(\ref{eq06}) one can also
 transfer these results to the corresponding $L$'s and $W$'s at some
 other renormalization scale different from $f_0\sqrt{\chi_R}$.
 Going to the conventional renormalization scale $\mu=4\pi f_0$ and
 multiplying by an overall factor $128\pi^2$ one obtains the values
 of $\alpha_k$ and $\omega_k$ shown in table~\ref{tab05}.

 Due to the unexpected smallness of the ${\cal O}(a)$ contributions
 it is interesting to try a linear fit of the valence quark mass
 dependences setting all ${\cal O}(a)$ terms to zero:
 $\eta_0=\eta_1=\eta_2=0$.
 This is a fit with eleven parameters because in the formulas $L_{S4W6}$
 is always multiplied by $\eta_S$.
 The result is a reasonable fit but the chi-square is by about 10\% 
 larger then in the case of $\eta_{0,1,2} \ne 0$.
 The best fit values of the main parameters are in this case:
 $\chi_R=36.1(1.0),\; \alpha_5=2.08(14),\; \alpha_{85}=0.502(48)$.

\begin{figure}
\begin{center}
\includegraphics[width=4.5cm]{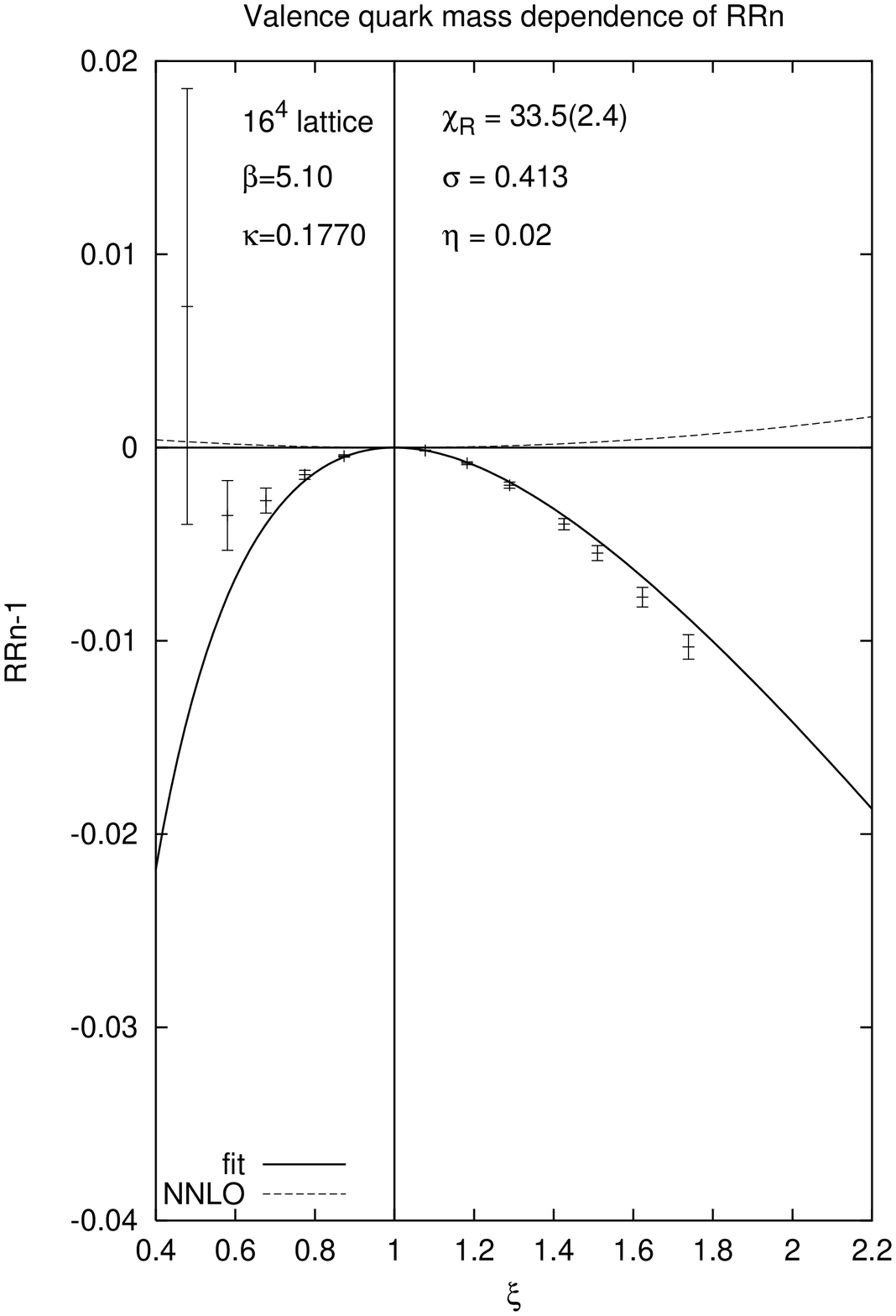}
\includegraphics[width=4.5cm]{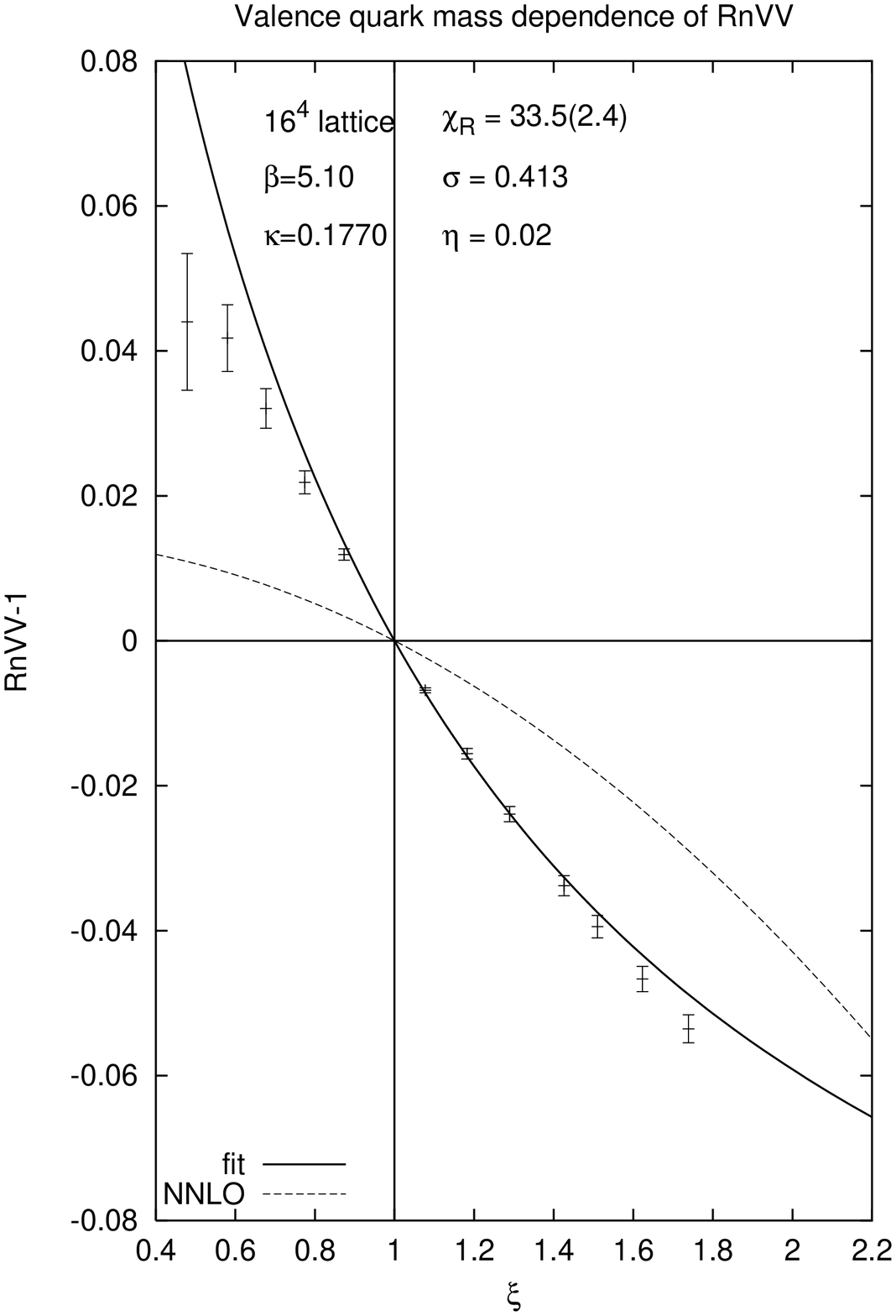}
\includegraphics[width=4.5cm]{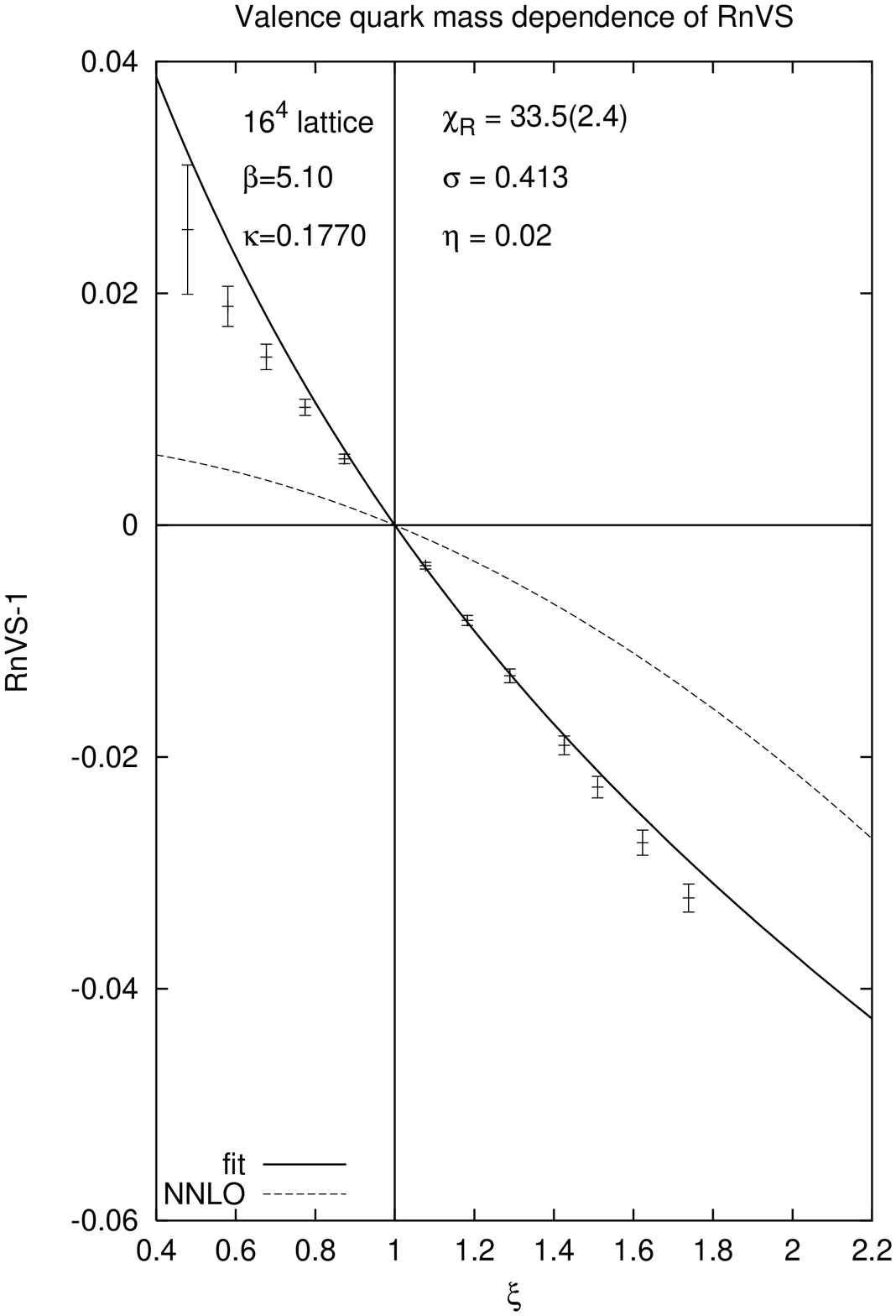}
\includegraphics[width=4.5cm]{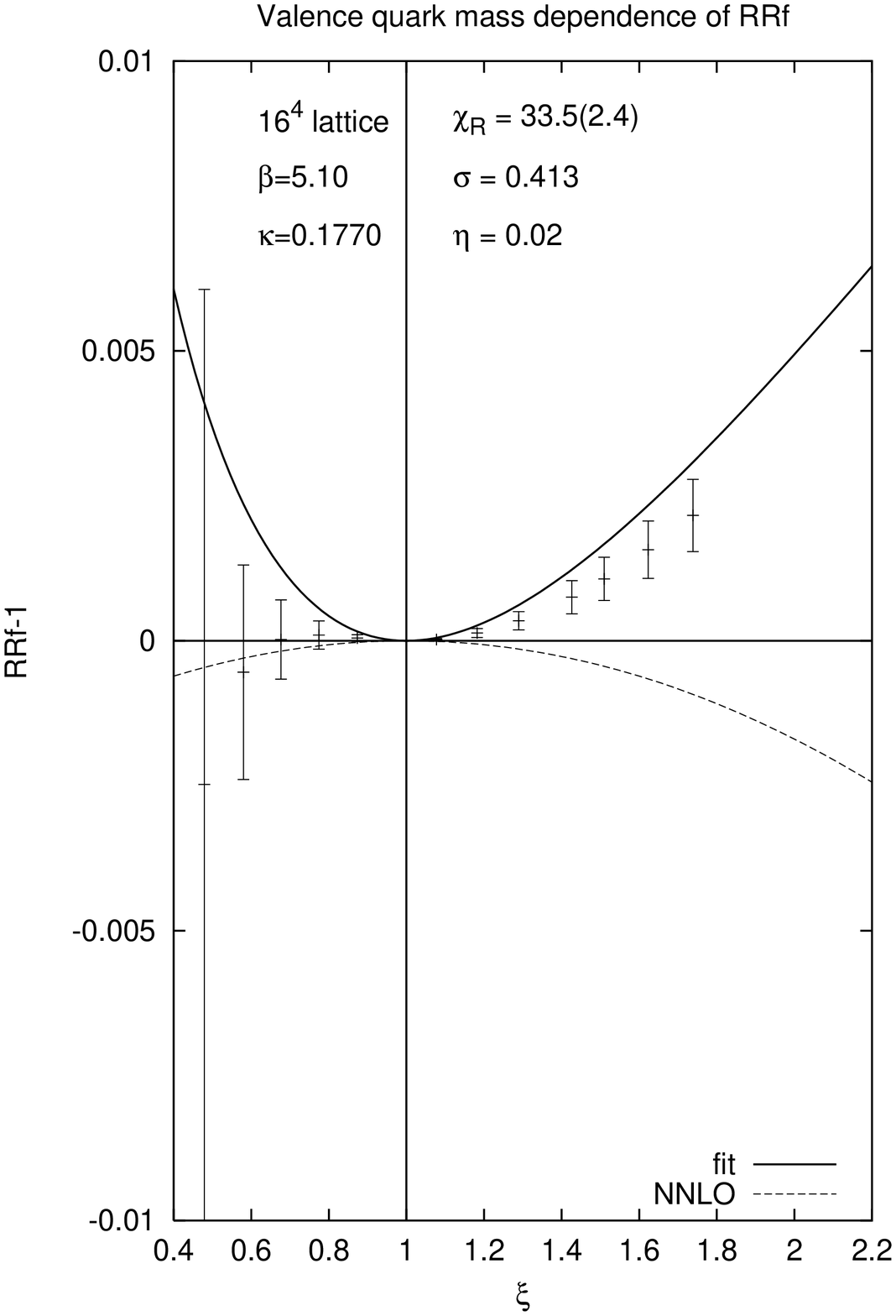}
\includegraphics[width=4.5cm]{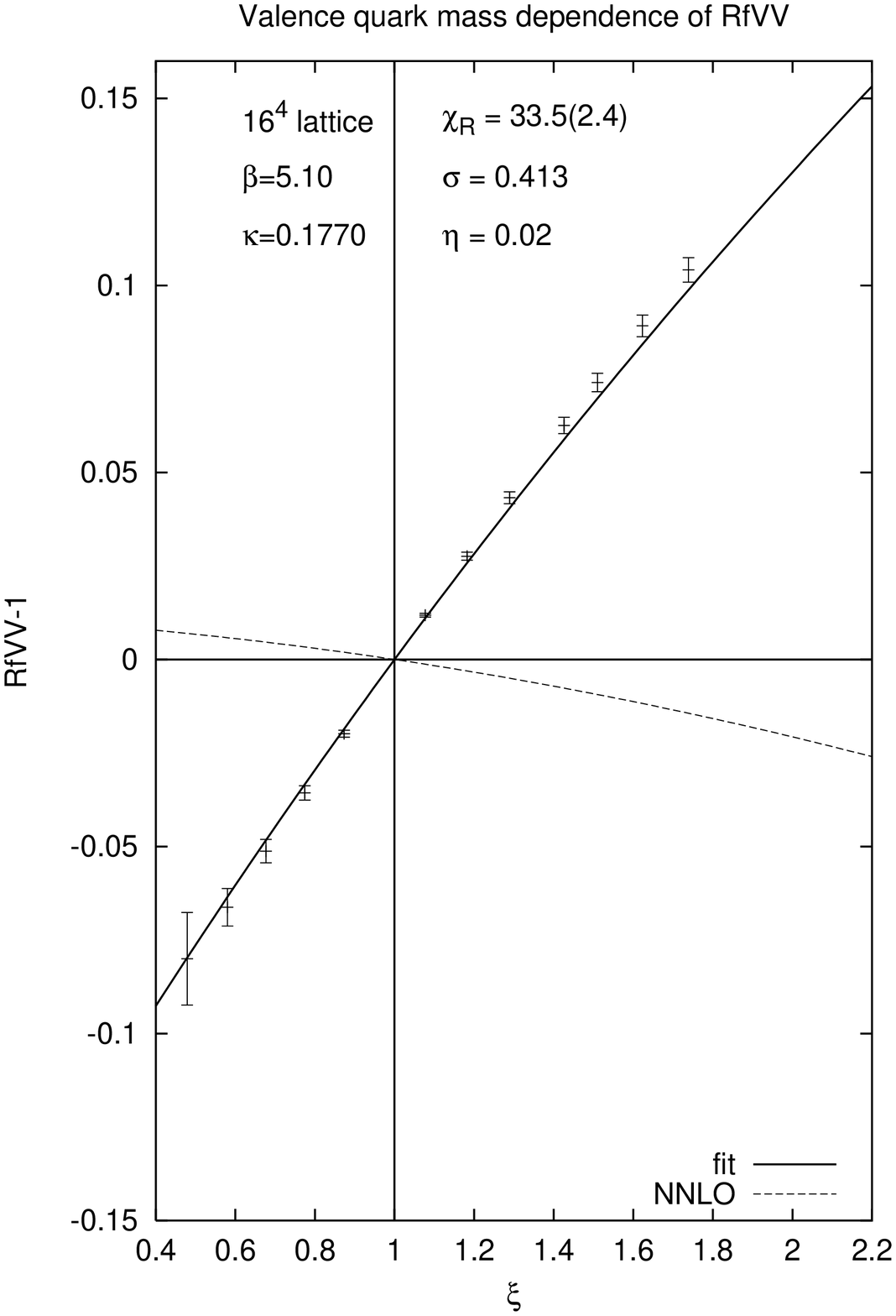}
\includegraphics[width=4.5cm]{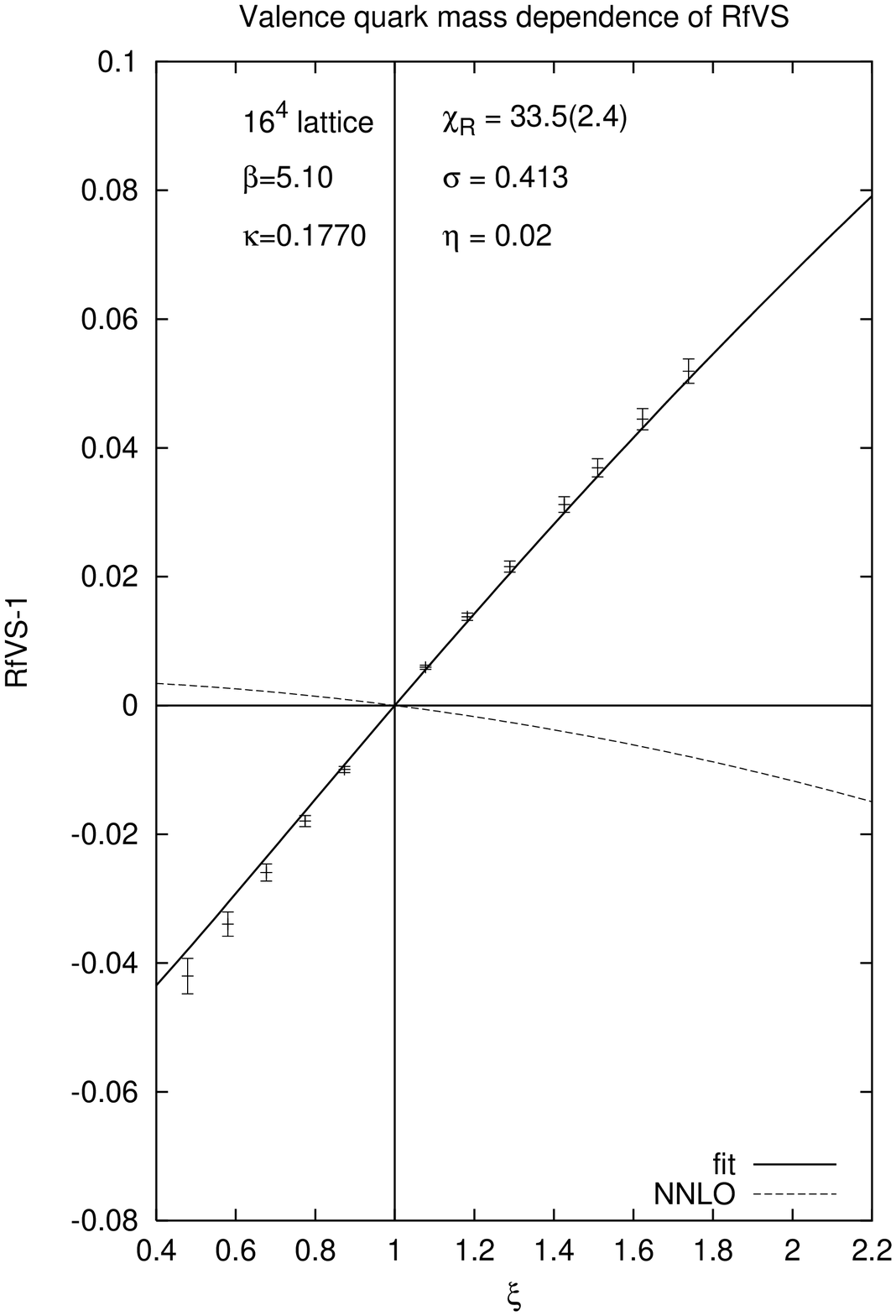}
\end{center}
\begin{center}
\parbox{12cm}{\caption{\label{fig03}\em
 NNLO tree-graph contribution at $\kappa_2=0.1770$ where the sea quark
 mass is given by $M_r \simeq 1$ (broken lines).
 The full lines represent the total fits shown also in figures
 \protect\ref{fig01}-\protect\ref{fig02} which are the sums of the
 continuum NLO, the ${\cal O}(a)$ and NNLO terms.}}
\end{center}
\end{figure}

 The NNLO tree-graph contributions are rather important especially at
 $\kappa=0.176$.
 From the point of view of the NLO formulas the situation becomes
 better at $\kappa=0.177$ but NNLO is still not negligible there: see
 figure~\ref{fig03}.
 (At $\kappa=0.1765$ we have, of course, an intermediate situation
 between $\kappa=0.176$ and $\kappa=0.177$.)
 In general, the NNLO contributions are more important in the ratios
 $R_{nVV}$ and $R_{nVS}$ than in $R_{fVV}$ and $R_{fVS}$.
 In fact, the ratios $R_{nVV}$ and $R_{nVS}$ at $\kappa=0.176$ are
 dominated by NNLO.
 The relative importance of NNLO terms is stronger for $\xi > 1$ than
 for $\xi < 1$.
 In the double ratios $RRn$ and $RRf$ the NNLO terms are relatively
 unimportant.

\subsection{Sea quark mass dependence}\label{sec4.2}

 The results from the fit of the valence quark mass dependence can also
 be used in the investigation of the sea quark mass dependence according
 to (\ref{eq22})-(\ref{eq23}).
 In particular, the values (and errors) of $\chi_R$ and $\eta_{0,1,2}$
 are relevant there.
 Besides these values and the known ratios of the sea quark masses
 $\sigma_{1,2}$ (see table~\ref{tab03}) two extra parameter pairs
 appear, namely, for $N_s=2$:
\be \label{eq31}
L_{R45} \equiv 2L_{R4}+L_{R5} \ ,  \hspace{2em}
W_{R45} \equiv 2W_{R4}+W_{R5}
\ee
 in (\ref{eq22}) and
\be \label{eq32}
L_{R6845} \equiv 4L_{R6}+2L_{R8}-2L_{R4}-L_{R5} \ ,  \hspace{1em}
W_{R6845} \equiv 4W_{R6}+2W_{R8}-2W_{R4}-W_{R5}
\ee
 in (\ref{eq23}).

\begin{figure}
\begin{center}
\includegraphics[width=9cm,angle=-90]{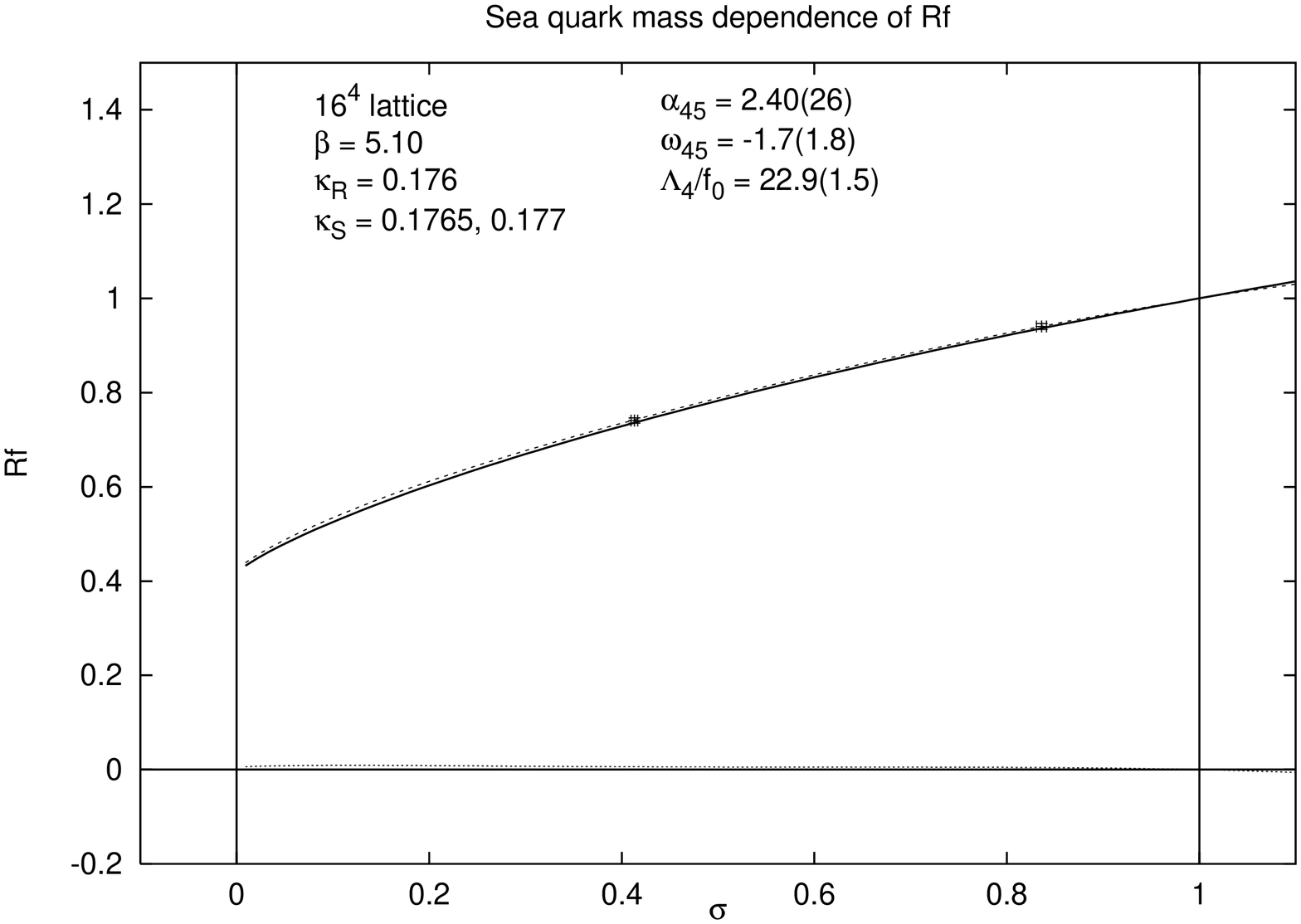}
\includegraphics[width=9cm,angle=-90]{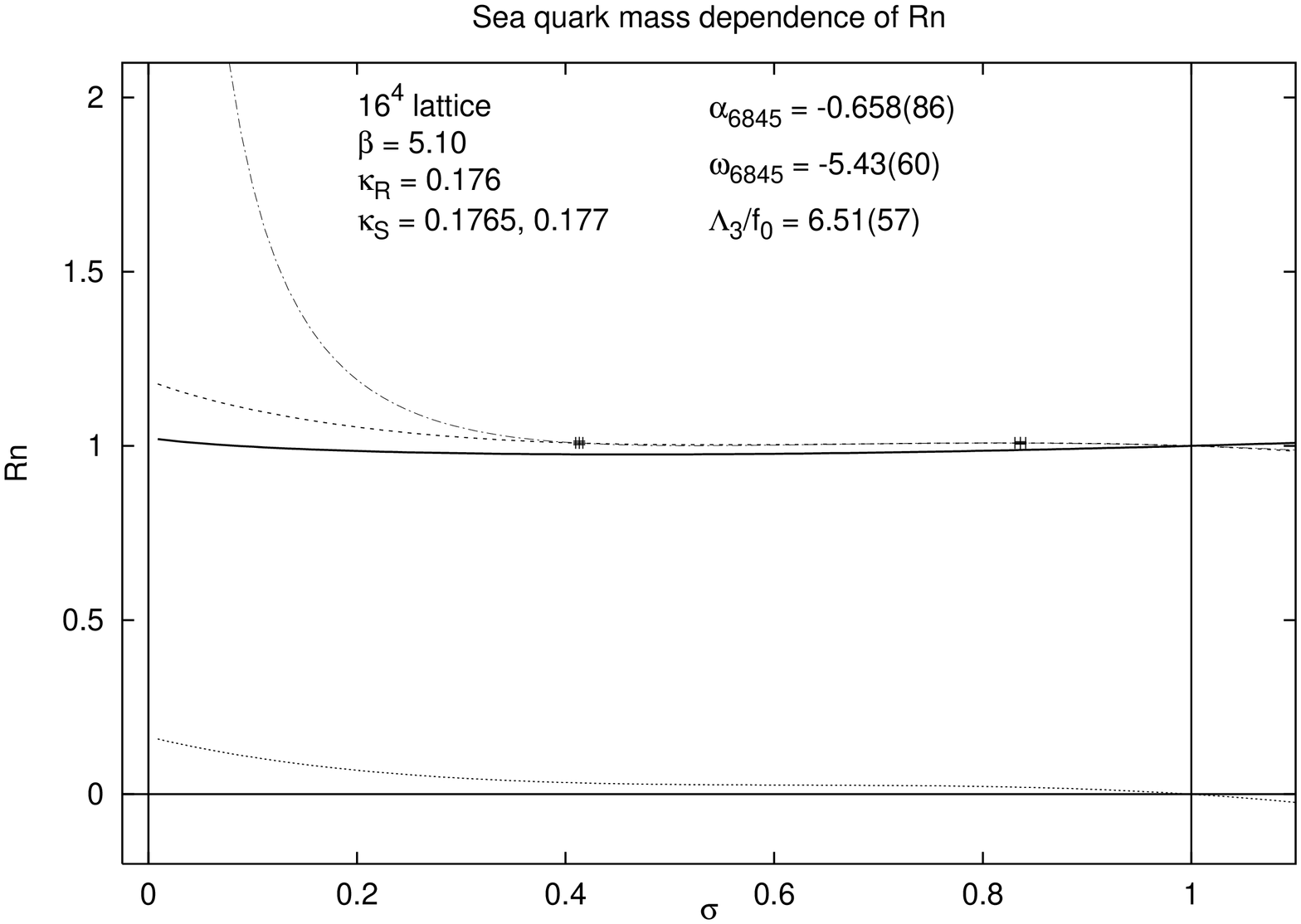}
\end{center}
\begin{center}
\parbox{12cm}{\caption{\label{fig04}\em
 $Rf_{SS}$ and $Rn_{SS}$.
 The full lines show the estimate of continuum contributions -- without
 ${\cal O}(a)$ terms.
 The broken lines near zero show the unphysical contributions
 (proportional to $\eta_S$).
 The other lines are different extrapolations of the measured values
 including ${\cal O}(a)$ terms (see text).}}
\end{center}
\end{figure}

 Since we only have three sea quark mass values and therefore two
 independent values of $Rf_{SS}$ and $Rn_{SS}$ a ``fit'' actually
 means solving for the four unknowns.
 The results are collected in table~\ref{tab06}.
 The corresponding values of the $\alpha$'s and $\omega$'s are
 contained in table~\ref{tab05}.
 In this table also the values of the {\em universal low energy scales}
 $\Lambda_{3,4}$ are given.
 (For the definitions see \cite{LEUTWYLER,DURR} or eq.~(10) in
 \cite{VALENCE}.)
 Once the parameters $L_{R45}$ and $L_{R6845}$ are known it is possible
 to extrapolate the continuum NLO curves (without the ${\cal O}(a)$
 contributions) for $Rf_{SS}$ and $Rn_{SS}$ to zero sea quark mass:
 see figure~\ref{fig04}.
 The values of these curves at $\sigma=0$ are also given in
 table~\ref{tab06}.
\begin{table}[ht]
\begin{center}
\parbox{12cm}{\caption{\label{tab06}\em
 Results for the parameters of the sea quark mass dependence.
 Quantities directly used in the fitting procedure are in bold face.}}
\end{center}
\begin{center}
\begin{tabular}{|c|r||c|r|}\hline
\bteq{L_{R45}}     & $4.34(28)\cdot10^{-3}$ &
   $Rf(\sigma=0)$            & $0.415(19)$
\\\hline
\bteq{W_{R45}}     & $1.1(1.4)\cdot10^{-3}$  &  &
\\\hline
\bteq{L_{R6845}}   & $-9.1(6.4)\cdot10^{-5}$ &
   $Rn(\sigma=0)$            & $1.025(17)$
\\\hline
\bteq{W_{R6845}}   & $-5.52(48)\cdot10^{-3}$  &  &
\\\hline
\end{tabular} 
\end{center}
\end{table}

 The extrapolation of the full measured ratios, including ${\cal O}(a)$
 contributions, requires an extrapolation of $\eta_S$ as a function of
 $\sigma$ which has, of course, a considerable uncertainty.
 The behavior of the extrapolated curve is especially sensitive
 to the assumed form of the $\eta_S$-extrapolation for $Rn_{SS}$ near
 zero.
 For instance, if the magnitude of the ${\cal O}(a)$ contribution given
 by $\rho_S=\eta_S\chi_S$ is finite at zero, which is reasonable to
 assume, then $Rn_{SS}=m_{SS}^2/(\sigma m_{RR}^2)$ has a $\sigma^{-1}$
 singularity near zero.
 This is a manifestation of the fact that different definitions of the
 ``critical line'' in the $(\beta,\kappa)$-plane, for instance by
 $m_\pi^2=0$ or $m_q^{PCAC}=0$, in general differ by lattice artifacts
 (in our case by ${\cal O}(a)$).
 If, however, $\eta_S=\rho_S/\chi_S$ would have a finite value at
 $\sigma=0$ then there would be no such singularity.
 The two extrapolations shown in the lower part of figure~\ref{fig04}
 are examples of these two cases.

 Concerning the results on the parameters obtained from the sea quark
 mass dependence (table~\ref{tab06} and the second half of
 table~\ref{tab05}) one has to remark that the assumption of a quark
 mass independent lattice spacing $a$ has an important effect on them.
 Assuming a quark mass independent Sommer scale parameter $r_0$ would
 change these results substantially.
 (There would be small changes in the first half of table~\ref{tab05}
 due to the somewhat different values of the quark mass ratios
 $\sigma_{1,2}$, too.)
 For instance, the values of $\Lambda_4/f_0$ and $\Lambda_3/f_0$
 would come out to be $16.1(1.1)$ and $30.4(2.9)$, respectively,
 instead of the values given in the tables.
 As it has been discussed above, the choice of a quark mass independent
 renormalization scheme requires a quark mass independent lattice
 spacing and is not consistent with a quark mass independent $r_0$
 \cite{BANGALORE,TSUKUBA}.
 Nevertheless, it is plausible that in the continuum limit and in the
 limit of very small sea quark masses $r_0/a$ becomes independent from
 the sea  quark mass and the differences between the values for constant
 $r_0$ and $a$ disappear.

\section{Summary and discussion}\label{sec5}

 The results obtained in this paper for the Gasser-Leutwyler constants
 (see tables~\ref{tab04}, \ref{tab05} and \ref{tab06}) can only be
 taken as estimates of the values in continuum.
 In order to deduce continuum values with controlled error estimates
 the left out lattice artifacts have to be removed by performing
 simulations at increasing $\beta$ values and extrapolating the results
 to $a=0$.
 Reasonable next steps would be to tune the lattice spacing to
 $a \simeq 0.13\, {\rm fm}$ on $24^3 \cdot 48$ and
 $a \simeq 0.10\, {\rm fm}$ on $32^3 \cdot 64$ lattices.
 This would require with the TSMB algorithm by a factor of about 10 and
 100 more computer time, respectively.
 Our calculations near $a \simeq 0.20\, {\rm fm}$ should be improved by
 going from $16^4$ to $16^3 \cdot 32$ lattices in order to improve the
 extraction of the physical quantities of interest.
 The number of sea quark masses considered should be increased to 5-6
 towards smaller values.
 This will decrease the overall statistical errors considerably.
 We hope to reach sea quark masses about $m_{sea} \simeq \frac{1}{6}m_s$
 on $16^3 \cdot 32$ lattices in the near future.

 General conclusions of the present work are:
\begin{itemize}
\item
 Compensating ${\cal O}(a)$ effects in the pseudo-Goldstone boson sector
 by introducing ${\cal O}(a)$ terms in the PQCh-Lagrangian itself is a
 viable alternative to the ${\cal O}(a)$-improvement of the lattice
 action.
 An extension to also treat ${\cal O}(a^2)$ effects in the
 PQCh-Lagrangian is possible \cite{BARUSH,AOKI} and has been partially
 taken into account also in the present paper.
\item
 The observed ${\cal O}(a)$ contributions in the pseudo-Goldstone boson
 sector are surprisingly small.
 The ratios of the ${\cal O}(a)$ parameters in the NLO PQCh-Lagrangian
 to the quark masses $\eta_S \equiv \rho_S/\chi_S$ are in our present
 range of quark masses
 ($\frac{1}{3}m_s \leq m_{sea} \leq \frac{2}{3}m_s$) at the few percent
 level.
\item
 Taking ratios of pion mass squares and pion decay constants at fixed
 gauge coupling and varying quark masses has the advantage that the
 $Z$-factors of multiplicative renormalization as well as all sorts of
 quark mass independent corrections cancel.
\item
 NNLO contributions in PQChPT are in our present sea quark mass range
 rather important.
 In fact, they are more important than the ${\cal O}(a)$ lattice
 artifacts.
 This introduces new parameters in the multi-parameter fits which
 makes the fitting procedure more difficult.
 The situation will be better at smaller sea quark masses where the
 importance of NNLO terms diminishes.
\end{itemize}

 The present results strengthen the observation already made in our
 previous paper \cite{VALENCE} that the expected behavior dictated by
 PQChPT sets in rather early -- at relatively large lattice spacings --
 once the quark masses are small enough.
 Our present cut-off $a^{-1} \simeq 1\, {\rm GeV}$ is already a
 ``high energy scale'' from the point of view of the pion dynamics.
 As a consequence, it seems to us that the numerical study of the
 pseudo-Goldstone boson sector of QCD is perhaps the easiest field for
 obtaining new quantitative results about hadron physics by lattice
 simulations.

\vspace*{1em}\noindent
{\large\bf Acknowledgments}

\noindent
 The computations were performed on the APEmille systems installed 
 at NIC Zeu\-then, the Cray T3E systems at NIC J\"ulich, the PC
 cluster at DESY Hamburg, and the Sun Fire SMP-Cluster
 at the Rechenzentrum - RWTH Aachen.
 Parts of the simulations were performed at the E\"otv\"os University
 parallel PC cluster supported by Hungarian Science Foundation grants
 OTKA-T349809/T37615.

 We thank Steve Sharpe for correspondence on the structure of the NNLO
 terms in the PQChPT formulas and Marteen Golterman and Oliver B\"ar
 for helpful discussions.
 We thankfully acknowledge the contributions of Claus Gebert in the
 early stages of this work.



\end{document}